# Evolution of bow-tie architectures in biology

Tamar Friedlander[1,2], Avraham E. Mayo[1], Tsvi Tlusty[3] and Uri Alon[1,*]

[1]Department of Molecular Cell Biology

Weizmann Institute of Science

Rehovot 76100, Israel.

[2]Institute of Science and Technology Austria,

Am Campus 1,

3400 Klosterneuburg, Austria.

[3]Simons Center for Systems Biology,

Institute for Advanced Study,

Einstein Dr., Princeton New Jersey 08540, USA.

[*] uri.alon@weizmann.ac.il

Classification: Biological Sciences: Systems Biology

Short title: Evolution of bow-tie architectures in biology

## Abstract

Bow-tie or hourglass structure is a common architectural feature found in biological and technological networks. A bow-tie in a multi-layered structure occurs when intermediate layers have much fewer components than the input and output layers. Examples include metabolism where a handful of building blocks mediate between multiple input nutrients and multiple output biomass components, and signaling networks where information from numerous receptor types passes through a small set of signaling pathways to regulate multiple output genes. Little is known, however, about how bow-tie architectures evolve. Here, we address the evolution of bow-tie architectures using simulations of multi-layered systems evolving to fulfill a given input-output goal. We find that bow-ties spontaneously evolve when two conditions are met: (i) the evolutionary goal is rank deficient, where the rank corresponds to the minimal number of input features on which the outputs depend, and (ii) The effects of mutations on interaction intensities between components are described by product rule – namely the mutated element is multiplied by a random number. Product-rule mutations are more biologically realistic than the commonly used sum-rule mutations that add a random number to the mutated element. These conditions robustly lead to bow-tie structures. The minimal width of the intermediate network layers (the waist or knot of the bow-tie) equals the rank of the evolutionary goal. These findings can help explain the presence of bow-ties in diverse biological systems, and can also be relevant for machine learning applications that employ multi-layered networks.

## Significance Statement

Many biological systems show bow-tie architecture: a huge number of inputs are converted to a small number of intermediates, which then fan out to generate a huge number of outputs. Examples are metabolic and signaling networks. However, there is no explanation of how bow-ties evolve in biology. Here, we find that bow-ties spontaneously evolve when the information in the evolutionary goal can be compressed, known in mathematics as a deficient goal rank. This rank is a number that determines the size of the narrowest part of the bow-tie in the network. This offers a first mechanism to understand a common architectural principle of biological systems, and a way to quantitate the rank of the goals under which they evolved.

# Introduction

Many natural and engineered systems show a bow-tie architecture (1, 2). A bow-tie (also termed hourglass) architecture is a feature of multi-layered networks in which the intermediate layer has significantly fewer components than the input and output layers. The intermediate layer is called the "waist" (3), "knot" (1) or "core" (4) of the bow-tie and in gene-regulatory networks the 'input-output' (5) or 'selector' gene (6). Bow-ties mean that the network is capable of processing a variety of inputs, converting them into a small set of universal building blocks and then reusing these building blocks to construct a wide range of outputs (see Fig. 1).

A bow-tie architecture is found for example in metabolic networks (1, 7–9), where the large range of nutrients consumed by the organism is decomposed into a very small set of basic molecules (such as the twelve precursors of central metabolism including pyruvate, G6P, F6P, PEP, AKG, ACCOA (1, 10)) from which the organism builds again all of its biomass including carbohydrates, nucleic acids and proteins. In the central dogma, a huge range of genes can be converted into their corresponding proteins via the same 20 amino acid intermediates. In mammalian signal transduction, a set of less than 10 pathways mediates information transfer between hundreds of possible input signals and the resulting expression changes in thousands of genes (10–13) - the same pathways are co-opted in different cell types to connect different inputs and outputs. The human visual system consists of multiple layers of signal processing, where hundreds of millions of photoreceptors in the retina fan in to only about one million ganglion cells (14) whose axons form the optic nerve. These in turn fan out to parallel processing pathways in the visual cortex that detect pattern, color, depth and movement (15)[1]. Some developmental gene regulatory networks have bow-tie structures in which a single intermediate gene ('input-output' or 'selector' gene) combines information from multiple patterning genes (the input layers) and then initiates a self-contained developmental program by regulating an array of output genes (5, 6). Studies of other biological signaling networks such as epidermal receptor signaling (16), GPCR signaling (17), signaling in both the innate (18, 19) and the adaptive immune system also documented bow-tie organizations (4, 20).

Objects manufactured by humans do not evolve in the biological sense, however the ongoing process of technological innovation is thought to have shared features a with evolution (21, 22). Many non-biological networks show bow-tie architectures as well. This includes the world wide web (23), internet protocols (3), production pipelines and some economic systems – see Table 1. Bow-ties in technology have, in a sense, evolved. For example, whereas in the past each machine had its own energy source (river for mill, fire for stove), in today's power grid a universal intermediate - 220V 50Hz AC electricity - connects multiple input energy sources (coal, oil, solar etc.) to multiple output appliances (1).

---

[1] One hypothesis is that the compression in the visual system is due to space limitations.

Bow-ties have been suggested to have functional implications. Bow-ties allow evolvability, because new inputs can be readily converted to new outputs, using the same well-tested intermediate processes (2). On the other hand, bow-ties are vulnerable to damage in the intermediate processes (1, 20). In developmental gene regulatory networks, modulated expression of the 'waist' ('input-output' or 'selector') gene can result in markedly different phenotypes. Thus it is thought that these 'waist' genes are hotspots for the evolution of novel phenotypes (5, 6). Once a bow-tie is established, it is hard to change its core components because changes to the bow-tie affect many processes at once (2, 3). Recently, Polouliakh et al. hypothesized that the narrow intermediate layer in signaling networks may serve to distinguish between different sets of inputs and assigns the correct set of outputs for each. As this intermediate layer is narrow compared to the number of inputs, different inputs are grouped together and share a common output response (17).

The prevalence of bow-tie architectures in biology beyond what one would expect if networks were generated at random, raises the question of how they evolved. In particular, one may ask whether there are evolutionary mechanisms that spontaneously give rise to bow-ties. This question is significant when considering the fact that most evolutionary simulations of multi-layered networks do not automatically give rise to bow-ties (3, 24). Generically, in fields as diverse as artificial neural networks (25) and evolution of biological networks, simulations result in highly connected networks with no bow-tie (26–32).

Bow-tie evolution in the context of internet protocol networks was studied by Akhshabi and Dovrolis (3). Their model assumed that the node connectivity monotonously decreases between layers, namely protocols at the input layer are general in terms of their function (have many connections each), and become more and more specific towards the output layer (in which they often have only a single connection). Bow-tie structures are then a direct outcome of this inhomogeneity in properties between layers. Such assumptions are relevant for technological applications, but are not relevant in the biological context. We thus sought a biologically plausible mechanism.

We were inspired by recent advances in understanding the evolution of a different feature that is common to various biological networks – modularity. Using simulations, several studies showed that evolution under modular goals with rules that tend to eliminate connections spontaneously lead to modular structure (24, 26, 33–35). This led us to ask whether one can find situations in which evolution spontaneously leads to bow-tie architectures.

Here, we study the evolution of bow-tie architecture using several simple models of multi-layered networks and biologically plausible evolutionary scenarios. We find that bow-ties evolve when two conditions are met: (i) the evolutionary goal has deficient rank; (ii) The effects of mutations on interaction intensities between components are described by a product rule – namely the mutated element is multiplied by a random number. Product-rule mutations are more biologically realistic than the commonly used sum-rule mutations that add a random number to the mutated element (36–42). For a detailed discussion of product-mutations, their biological relevance and their evolutionary effect, the reader is referred to an earlier work (24). We further show that the narrowest possible waist in the bow-tie is equal to the rank of the goal. We demonstrate this in simulations of evolution in several model systems.

**Results**

**Simulations of multi-layered network models evolving towards input-output goals**

We begin with the simplest model of a multi-layered network, namely a linear model, and then turn to non-linear models. Related models have been useful for understanding multi-layered biological networks in diverse contexts (28, 33, 41, 43–52). We study feed-forward networks, namely connections are only possible between a node and any node at the next layer. Connections within layers or backward connections are not allowed. In the model, the interaction intensities between every two consecutive network layers are represented by a matrix. Thus, a network is described by $L$ matrices $A^{(1)}, A^{(2)} \ldots A^{(L)}$, where the number of layers (including input layer) is $L+1$ (Fig. 1). The input-output relationship of the network is given by the product of these matrices, which represents the total transfer of signals from the first (input) to the last (output) layer.

We evolve these networks to fulfill a goal - a prescribed input-output relation given by a matrix $G$. The dimension of the goal matrix $G$ corresponds to the number of system inputs and outputs $D_{output} \times D_{input}$, The fitness of a network equals the distance between $G$ and the product of the matrices $F = -\|A^{(L)}A^{(L-1)} \ldots A^{(1)} - G\|$. Note that each goal can be satisfied by an infinite number of matrix combinations. For example, if $G$ is the unit matrix $G = I$ and $L = 2$, all pairs of matrices that are inverse to each other $A^{(2)} = [A^{(1)}]^{-1}$ satisfy the goal, because $A^{(2)} \times A^{(1)} = I$.

To evolve the networks, we use a standard evolutionary simulation framework (32, 53–55). Briefly, the simulation starts with a population of $N$ networks, each described by $L$ matrices. At each generation the networks are duplicated and mutated with some probability resulting in modified interaction intensities. A mutation means a change to an element of one of the matrices. Fitness is then evaluated for each structure in comparison to a goal. $N$ individuals are then selected to form the next generation of the population, such that fitter individuals are more likely to be selected. This process is repeated, until high fitness evolves (see Methods section for more details).

Guided by studies on the evolution of modularity, we tested evolutionary situations which are biased to reduce or eliminate interactions. Such a mechanism is a product-rule mutation scheme in which elements of the matrices are multiplied by a random number drawn from a normal distribution $N(1, \sigma)$ (as opposed to a sum-rule mutation where a random number is added instead of multiplied) (24). Product-rule mutations are a more realistic description of the way that DNA mutations affect biochemical parameters than sum-rule mutations (36–42). Biological mutations are more likely to decrease existing interactions than to create novel ones that did not exist before (56–58). This property is captured by product-rule mutations (but not by sum-rule) (24). With this mechanism, evolution finds networks satisfying the goal which are highly sparse - that is, networks with a small number of significant interactions (24, 41). As controls, we also simulated evolution with sum-rule mutations (in which a random number is added to matrix elements).

**Bow-tie architectures evolve when the goal is rank deficient**

We tested evolution towards goals described by matrices with different ranks. The rank $r$ is the number of linearly independent rows in the matrix. The rank of the goal matrix is full, if all rows of the matrix are independent. If some of the rows are dependent, the matrix has deficient rank – a rank smaller than the full rank. Deficient rank means that the input-output transformation maps inputs to a limited subspace of outputs, of dimension $r$. Below, we discuss the implications of this concept also for nonlinear systems. As an example of a $3\times 3$ matrix with rank $r=1$ consider the following matrix whose two last rows are given by a constant multiplying the first row:

$$G = \begin{pmatrix} v_1 & v_2 & v_3 \\ \alpha v_1 & \alpha v_2 & \alpha v_3 \\ \beta v_1 & \beta v_2 & \beta v_3 \end{pmatrix}.$$

The reason for testing rank-deficient matrices arises from the observation that they can be decomposed into a product of (generally non-square) matrices $A^{(L)}{}_{D_{output} \times D_{L-1}} \times \cdots \times A^{(2)}{}_{D_2 \times D_1} \times A^{(1)}{}_{D_1 \times D_{input}} = G_{D_{output} \times D_{input}}$, whose smallest dimension equals the rank of the goal, namely $\min_i(D_i) = r$. Because the rank of the goal matrix is smaller than its dimension $r < D$, this decomposition is equivalent to a narrow waist, whose width is equal to the goal matrix rank $r$. As a simple example consider the $3 \times 3$ goal above. It is decomposable into a product of a column vector by a row vector:

$$G = \begin{pmatrix} v_1 & v_2 & v_3 \\ \alpha v_1 & \alpha v_2 & \alpha v_3 \\ \beta v_1 & \beta v_2 & \beta v_3 \end{pmatrix} = \underbrace{\begin{pmatrix} 1 \\ \alpha \\ \beta \end{pmatrix}}_{A^{(2)}} \underbrace{\begin{pmatrix} v_1 & v_2 & v_3 \end{pmatrix}}_{A^{(1)}}.$$

This decomposition represents a 3-layer network, whose intermediate layer has only one active node – namely has a bow-tie structure (see left scheme in Fig. 1B).

More generally, let $G_{D_{\text{output}} \times D_{\text{input}}}$ be a goal matrix with rank $r = \text{rank}(G)$. Let there be a decomposition of $G_{D_{\text{output}} \times D_{\text{input}}}$ into a product of $L$ matrices $G = A^{(L)} A^{(L-1)} \ldots A^{(1)}$. This $L$ matrix decomposition is a representation of a network that has $L+1$ layers of nodes, whereas each matrix represents the interaction intensities between all nodes in two adjacent layers. If and only if $G$ has deficient rank, it can be decomposed into a product of matrices having dimensions smaller than the goal dimensions. This means that the matrices represent a network with intermediate layers that consist of fewer active nodes than the number of inputs and outputs to the network. Otherwise, if the matrix has full rank, each layer must have a number of active nodes which is at least as large as the rank, making a bow-tie impossible in a case of full rank. This argument follows from the fact that matrix multiplication cannot increase rank, i.e. $\text{rank}(AB) \leq \min(\text{rank}(A), \text{rank}(B))$ for any two matrices $A$ and $B$ (59).

The narrowest layer in the network is termed the waist (3). The waist is narrow because of the low rank that allows compressing the inputs down to fewer nodes, and then computing the outputs based on those nodes. While in principle such deficient decompositions exist for every rank-deficient matrix, they constitute only a small fraction of all possible decompositions.

The existence of bow-tie representations still does not guarantee that evolution can find them among the infinite number of possible solutions. We therefore evolved networks towards rank-deficient goals, with and without the product-rule mutation scheme described above. We studied goals of dimension $D = 6\text{-}8$ consisting of $L = 4\text{-}6$ matrices, tested 4-8 different goals for each dimension, and evolved networks towards each goal in 100-3000 repeated simulations, each starting from different random initial conditions. We found that deficient-rank goal together with product-rule mutations gave rise to networks that satisfy the goal and show bow-tie architectures. Full rank goals always led to evolved networks that satisfied the goal but had no bow-tie architecture at all, namely all layers had exactly the same number of nodes as the input and output layers. Rank-deficient goals with a different mutational scheme (sum rule mutations) could sometimes lead to architectures in which intermediate layers had fewer nodes than the input and output layers, but these were mostly not as narrow as the goal rank, regardless of run-time.

For example, consider a network with 5 layers of nodes ($L = 4$ matrices of interaction layers), each consisting of 6 nodes ($D_{\text{input}} = D_{\text{output}} = D = 6$). We simulated their evolution towards goals of different ranks between 1 and full rank ($r = 1\text{-}6$). We repeated the simulation 700 times for each goal starting from different random matrix initial conditions. We then analyzed the number of active nodes at each layer. Since in a numerical simulation we do not obtain exact zeroes, but rather very small values, we defined active nodes as nodes who, if removed (incoming and outgoing interactions of the node set to zero) have a larger than 0.1% relative effect on fitness (see Methods).

We find that the number of active nodes is smallest on average at the middle layer (Fig. 3). The number of active nodes at this waist is most often equal to the rank of the goal (Fig. 2), and never lower than this rank. The first and last layers are constrained to have exactly $D$ active nodes by the definition of the problem. Not all runs however reached the most parsimonious configuration possible: they sometimes (~20%, see Table S1) showed more active nodes than the rank of the goal. For comparison, if the mutational mechanism is not biased to decrease interactions (i.e. sum-mutations) 94%-97% of the runs ended with mid-layer which had more elements than the rank of the goal (Table S1). We show a representative example of a network configuration obtained in simulation in Fig. 1C.

We tested the sensitivity of this mechanism for bow-tie evolution to model parameters. A bow- tie was obtained under a wide range of values of selection intensity, mutation size, mutation rate and population size that spanned 1.3 decades (mutation rate) to 2 decades (mutation size) (Fig. 6, see SI Appendix for more details). We also tested the sensitivity of the structure obtained to the evolutionary goal by comparing simulation results with different goals having the same rank. We find that the location and width of the waist are insensitive to the choice of the goal (see Fig. S9 in SI Appendix).

We tested the location of the waist in simulations of multi-layered networks with equal number of inputs and outputs ($L=4, D=6; L=6, D=8$). While in principle the waist could reside at any layer between the input and output layers, in practice, it falls most often in the middle layer. Intuitively, this can be explained by symmetry considerations: The mutational mechanism works uniformly on all layers to eliminate connections. While the dimensions of the goal matrix constrain the number of active nodes at the network boundary layers (input and output), connections near the middle layer are least "protected" and thus mostly prone to removal, resulting in the network waist being on average in the middle layer.

**"Noisy rank" also leads to bow-tie architecture**

We next tested the effect of adding random noise to the rank-deficient goal matrix. This produced matrices that are 'almost rank deficient': full rank, but with some of the eigenvalues close to zero. The noise strength is given by the difference between the norms of the noisy and clean goals divided by the norm of the clean goal (see Methods). We find that for noise strength up to about 1%, bow-tie architecture with middle layers whose width equals the goal rank were reached in most simulation runs, just as in the absence of noise. Thus, our evolutionary simulation is robust to small perturbations to exactly rank-deficient goals – see Fig. 4 for illustration and compare to Fig. 2 with no noise. The median waist size increased above the clean rank when noise intensity increased above 1% (see Figs. S14-S15 for the dependence of bow-tie on the noise level). We also find that adding noise to the evolved matrices, simulating biological noise in network components, does not affect the results as long as noise is not too large (data not shown).

**Nonlinear information transmission**

Finally, we asked whether the present mechanism would apply in a nonlinear network model. While goal rank is a straightforward measure of dimensionality in linear systems, the concept of rank is more elusive when it comes to nonlinear systems. Yet, one can intuitively think that a similar concept could exist there too. To test this hypothesis we employed a well-studied problem of image analysis using a widely used molde: percepton nonlinear neural networks (60, 61). In this problem each node integrates over weighted inputs and produces an output which is passed through a non-linear transfer function, $u^{(l+1)} = f(A^{(l)}u^{(l)} - T^{(l+1)})$, where $A^{(l)}$ and $T^{(l)}$ are the weights matrix and corresponding set of thresholds in the $l$-th layer, and $u^{(l)}$ is the set of inputs propagated from the previous layer (see Methods).

We evolved the networks towards a goal of identifying features in a $2 \times 2$ retina with Boolean pixel values ($D_2 = 4$ inputs) (Fig. 5A). Low dimensionality was achieved by defining as a goal four outputs that depend only on two features of the image. The four required Boolean outputs were: (a) at least one pixel in the left retina column, (b) at least one pixel in the right column, (c) pixels in both left and right columns, (d) pixel(s) in the left or in the right columns. These four outputs can be fully represented by only two features: (a) and (b), making the 4-dimensional input space redundant. Thus, the effective "rank" here is $r = 2$.

Simulations evolving the weights $\{A_{ij}^{(l)}\}$ and thresholds $\{T_i^{(l)}\}$ values using product-rule mutations that led to perfect solutions (fitness of less than $10^{-4}$ from the optimum) mostly had a narrow waist: one of the intermediate layers had only two active nodes in 75% of the runs (see Table S1 in SI Appendix). For comparison, simulations with a mutation rule that was not biased to eliminate interactions (sum-rule) were much less likely to lead to networks with a narrow waist (this was observed in only 45% of runs). Detailed statistics over 500 runs of network structures obtained with either mutational scheme is presented in Fig. S12.

**Discussion**

We studied the evolution of bow-ties in layered networks. We find that bow-ties evolve spontaneously when two conditions are met: the goal has deficient rank and the effect of mutations on interactions is well-approximated by a product-rule. The size of the narrowest layer - the waist of the bow-tie - is bounded from below by the rank of the goal. We find the evolution of narrow waists in a wide range of evolutionary parameters, in both linear and nonlinear multi-layered network models.

The concept of rank is defined clearly in the case of matrix-like goals and linear transfer functions. In more complex situations, such as the nonlinear retina problem and gene regulatory networks, the rank corresponds to the minimal number of input features on which the outputs depend. One may hypothesize that in the case of probabilistic time dependent signaling in cells and nervous systems, rank may be related to the information theory measure of information source entropy. This is the minimal number of bits which is sufficient to encode the source (62). A natural information source ("input") - such as biological signals - is often redundant. Its compression (source coding) can shorten the description length while still preserving all the necessary information ("waist"). In analogy to the goal rank, the shortest possible description equals the source entropy. The present results can supply an operational definition of the goal rank in a layered nonlinear system - the minimal evolved waist under the present assumptions (63–69).

The present findings suggest that the proposed evolutionary mechanism leading to bow-ties is robust to small perturbations to exactly rank-deficient goals. This ability to distinguish dominant from negligible matrix eigenvalues is useful in analysis of high-dimensional datasets. It is proposed that bow-ties may be a natural dimensionality reduction mechanism of biological systems. For example, a recent study characterized the response repertoire of apoptotic signaling and found that the seemingly large space of inputs reduces to only a handful of significant 'biological axes' (70).

Bow-tie structures are also common in multi-layered artificial neural networks used for classification and dimensionality reduction problems. While there are parallels in the functional role of bow-ties there with the biological bow-ties which are the focus of this study, these neural networks are designed a priori to have this structure. Multi-layered neural networks often use an intermediate (hidden) layer whose number of nodes is smaller than the number of input and output nodes (25, 71). There, the role of the hidden layer is to capture the significant features of the inputs. The favorable usage of bow-tie structures in neural networks suggests that often the number of important features is lower than the number of inputs (60). The transformation between input and hidden layer was shown to map the data into a space in which discrimination is easier (72, 73). Recently, detailed mappings of signaling networks suggested a similar role for small molecules in GPCR signaling of mammalian immune cells (17). There, a narrow intermediate network layer works as a classifier that can match the corresponding output to a large set of different inputs.

Taken together, our results suggest a first mechanism for the evolution of bow-tie architectures in biology and a way to quantitate the rank of the evolutionary goals under which they evolved.

**Materials and Methods:**

**Evolutionary simulation**

Simulation was written in Matlab (53–55). All source codes, data and analysis scripts are freely available in a permanent online archive at XXX (upon paper acceptance). We initialized the population of matrices by drawing their $N \cdot LD^2$ terms from a uniform distribution. Population size was set to $N = 100$. Each "individual" consists of a set of $L$ matrices. In each generation the population was duplicated. One of the copies was kept intact, and elements of the other copy had a probability $p$ to be mutated – as we explain below. Fitness of each of the $2N$ individuals was evaluated by $F = -\|A^{(L)}A^{(L-1)}...A^{(1)} - G\|$, where $\|\cdot\|$ denotes the sum of squares of elements (74). The best possible fitness is zero, achieved if $A^{(L)}A^{(L-1)}...A^{(1)} = G$ exactly. Otherwise, fitness values are negative. In the figures we show the absolute value of mean population fitness, which is the distance from maximal fitness. We constructed the goal matrices from combinations of zero and 10 terms. We tested goals of different ranks and different internal structures and found no sensitivity for goal details other than its rank (see SI Appendix). $N$ individuals are selected out of the $2N$ population of original and mutated ones, based on their fitness (see below). This mutation–selection process was repeated until the simulation stopping condition was satisfied (either a preset number of generations or when mean population fitness was within 0.01 of the optimum).

**Mutation**: We mutated individual elements in the matrix. We set mutation rate such that on average 20% of the population members were mutated at each generation, so the probability of each matrix element to be mutated was $\sim \frac{0.2}{LD^2}$. This relatively low mutation rate enables beneficial mutants to reproduce on average at least 5 generations before an additional mutation occurs.

We randomly picked the matrix elements to be mutated. Mutation values were drawn from a Gaussian distribution (unless otherwise stated). The mutated matrix element was then multiplied by the random number: $A_{ij}^{(l)} \rightarrow A_{ij}^{(l)} \cdot \mathrm{N}(1, \sigma)$. In simulations we used $\sigma$ in the range 0.01-1. Maximal achievable fitness and the time-scale to convergence depend on the mutation frequency and size, as demonstrated in our sensitivity test (see SI Appendix).

**Selection methods**: We used tournament selection with group size $s = 4$ (see (55) chap. 9). In a previous work we tested 2 other selection methods (truncation-selection (elite) (53) and proportionate reproduction with Boltzmann-like scaling (41, 50, 75)) and found that all three methods gave qualitatively very similar results with only a difference in time scales.

**Noisy goals**

In order to test the effect of noisy rank we added a low amount of noise to the goals used in the previous simulations. We used goals with ranks 1, 2 and 3 whose terms were either 10 or 0 and then added a uniformly distributed noise in the range [0, 0.1] ($\sigma$ = 0.029). We define the noise level as the absolute value of the difference between the norms of the noisy and clean goals divided by the norm of the clean goal: $\left|\frac{\|G^*\|-\|G\|}{\|G\|}\right|$, where $G$ is the 'clean goal' and $G^*$ is the noisy one. As norm we took the sum of squares of all matrix terms. At every repeat of the simulation we added a different noise realization with the same statistics. The noise (and thus the evolutionary goal) was fixed throughout any given run. The noise intensity was calculated separately for each run. The values presented are averaged over all runs considered in the analysis.

**Data analysis:**

Repeated simulations were run using the same parameters, where at every single run the Matlab random seed was initialized to a different value. Consequently, each run starts from different initial conditions and uses different mutational realizations. In the analysis, we checked whether the runs converged. Only runs that gave results within 0.01 from the optimum were considered in the analysis. We then analyzed in each run the number of active nodes in the layer (see below). In the figures we show either the median number or histogram of active nodes per layer.

**Active nodes:** To calculate the number of active nodes in a layer, we eliminated each node at a time, by equating to zero all its input and output interactions. For example to eliminate the $k$-th node in layer $l+1$ we set $A^{(l)}_{k,*} = A^{(l+1)}_{*,k} = 0$, leaving all other terms intact. We then calculate the fitness value of the modified network $\tilde{F}$ and define the difference compared to the original fitness value $F$: $\Delta F = |F - \tilde{F}|$. We compare $\Delta F / F$ between all nodes located at the same layer. A node whose relative effect on fitness is less than 0.1% is considered inactive.

**Retina problem**

We tested the evolution of bow-tie network in this non-linear problem which resembles standard neural network studies (34, 60, 76). We defined a problem with 4 inputs and 4 outputs and 2 internal processing layers consisting of 4 nodes each. The inputs represent a 4-pixel retina, where each pixel could be either black or white, as described in the results section.

The evolutionary simulation followed a similar procedure to the linear problem described above. Mutation, selection, and data analysis methods were similar to the ones used in the linear problem as described above. The main difference is that the output of each layer was not a linear function of the inputs as before, but rather a non-linear function $u^{(l+1)} = f(A^{(l)}u^{(l)} - T^{(l+1)})$, where $A^{(l)}$ and $T^{(l)}$ are the weights matrix and corresponding set of thresholds in the $l$-th layer correspondingly, and $u^{(l)}$ is the set of inputs propagated from the previous layer. The non-linear transfer function $f$ was rescaled to range between 0 and 1, $f(x) = (1 + \tanh(x))/2$. The result of this computation is fed to the next layer until the last (output) layer is reached. In non-linear systems the evolutionary goal cannot be described by a single goal matrix as in the linear case. Rather, it is defined by pairs of input /output relations. The evolutionary simulation tested all possible inputs simultaneously, and evolved the network parameters to provide the correct output in each case. Inputs and outputs were encoded by Boolean vectors. Internal layer calculation used continuous values, but simulations could reach very high precision ($\leq 10^{-10}$ from the optimum). Simulations were run for $10^4$ generations. Only runs that reached fitness within $10^{-4}$ of the optimum were considered in the analysis. The fitness was defined as the difference between the network output and the desired output, in similarity to the linear model and then averaged over all possible input/output pairs.

The retina simulation was written in Wolfram Mathematica. We initialized the population of matrices and corresponding thresholds by drawing their $N \times LD(D+1)$ terms from a uniform distribution in the range [-2,+2]. Population size was set as $N = 100$. In each generation the population was duplicated. One of the copies was kept intact, and elements of the other copy had a per-term probability $p = 0.2$ to be mutated. The mutation was implemented through multiplying the mutated term by a random number drawn from a normal distribution with mean 1 and std 0.5 (thus a probability of about $q = 0.02$ to change sign).

To determine active nodes in this case, we begin by setting each weight to zero in its turn $A^{(l)}_{ij} = 0$ leaving all other terms intact. This procedure was not applied to the threshold values, $T^{(l)}_i$, because a node may be left in the network, even if no inputs are propagated through it from an upper layer. In these cases the role of such a node is to introduce a constant bias set by its threshold. We then calculate the fitness value of the modified network $\tilde{F}$ and define the difference compared to the original fitness value: $\Delta F = |F - \tilde{F}|$. We compare $DF/F$ between all weights located at the same layer. A network interaction whose relative effect on fitness is less than $10^{-4}$ was set to zero. A node whose entire set of outgoing weights was set to zero was considered inactive.

## Acknowledgements

We thank Tiago Paixao, Georg Rieckh, Hila Sheftel, Pablo Szekely, Gašper Tkačik and Marcin Zagorski for critical reading of the manuscript; and Katarina Bodova, Christoph Lampert, Virginie Orgogozo, Vitaly Shaferman and Richard Wallbank for useful discussions. The research leading to these results received funding from the Israel Science Foundation and the European Research Council under the European Union's Seventh Framework Programme (FP7/2007-2013) /ERC Grant agreement n° 249919. U. A. is the incumbent of the Abisch-Frenkel Professorial Chair. T.F. acknowledges partial funding from the People Programme (Marie Curie Actions) of the European Union's Seventh Framework Programme (FP7/2007-2013) under REA grant agreement n° 291734.

**Figure Captions**

**Fig. 1 – Model description. (A)** Bow-tie in a multi-layered network means that the network is capable of processing many different inputs, by converting them into a small set of universal building blocks and then re-using these building blocks to construct a wide range of outputs. **(B) Multi-layered networks are represented by interaction intensities between components:** Our model represents a multi-layered information transmission network, by the values of interaction intensities between nodes in consecutive layers. In this schematic figure we illustrate networks with 3 layers of nodes, connected by $L$ =2 layers of interactions. It is convenient to recapitulate these interactions by $L$ =2 matrices, where the $A_{ij}$ term in the $l$ -th matrix represents the interaction between the $j$-th component in node layer $l$ to the $i$-th component in node layer $l+1$. Node layer 1 is the input signal, and node layer $L+1$ is the output. In general, every node could be connected to every node in the next layer – as in the rightmost scheme. A bow-tie is a situation in which in one or more of the middle layers some nodes are disconnected from the rest of the network. This forms a narrow layer, termed "waist" – as exemplified in the left and middle schemes. A bow-tie architecture is captured by interaction matrices in which some rows/columns are zero. The number of non-zero rows/columns corresponds to the width of the waist layer. **(C) An example of bow-tie networks (simulation results):** An example of simulation results with $L=4$ interaction layers (5 node layers), demonstrating a bow-tie of width 1 at the middle layer. The network structure is shown on the left (only active nodes shown) and the interaction intensities are shown on the right using a color code (white – no interaction, black – strong interaction).

**Fig. 2 – Product-rule mutations and goal which is not full rank can lead to bow-tie architecture.** We show simulation results of networks with $L=4$ (5 layers of nodes) and 6 nodes in each layer ($D=6$). We performed 4 different sets of repeated simulations with goals of different ranks =1,2,3 or 6. We illustrate the histograms of layer width for each set of runs. Each column in this figure shows simulation results for a different goal, and each row shows a different network layer. The number of active nodes in middle layers varies depending on the goal. The minimal number of nodes in intermediate layers ("waist") is bounded from below by the rank of the goal. The waist width could be higher than the rank, because not all runs reach the most minimal configuration, but it cannot be lower. For example, it can be as low as 1 if the goal rank equals 1 (left column), but it is always 6 if the goal is full rank, demonstrating that no bow-tie can evolve with a full-rank goal. Simulation parameters: 3000 repeats for rank 1 and 2, 1500 repeats for rank 3 and 700 repeats for rank 6. Only runs that reached a fitness value less than 0.01 from the optimum were considered in the analysis. Product mutations were drawn from a Gaussian distribution with $\sigma=0.1$, element-wise mutation rate $p=0.05/D^2$, tournament selection with $s=4$.

**Fig. 3 – The waist is most likely to evolve in the middle layer (for equal number of inputs and outputs). Top:** Median number of nodes at each layer. Different curves represent results for goals of different ranks. Due to symmetry considerations, the waist is most likely to evolve in the middle layer of nodes. Results refer to the same simulations as in the previous figure. Estimation of error in median calculation by bootstrapping resulted in negligible error. **Bottom:** examples of possible network structures evolved with goals having different ranks 1,2,3 and 6, illustrating how the width of waist depends on the goal rank.

**Fig. 4 – Bow-tie evolves even if the goal is only approximately of deficient rank.** We show simulation results when the goal consisted of a matrix of deficient rank (1, 2 or 3) to which some level of noise was added (see Methods), so mathematically speaking goals had full rank, such that some of the eigenvalues were relatively small. Remarkably, here too a bow-tie architecture evolved, however the width of the waist was not as narrow as if the goal had exact noiseless

deficient rank (compare to Fig. 2). For each goal rank we calculated layer activity statistics based on 1500 different runs (each having a different goal, but with same noise statistics). Noise level here was 1% (averaged over all runs analyzed) for all ranks. This result demonstrates that the evolutionary process can expose a deficient goal rank even when noise is added, as is expected to be the case in realistic systems. Other parameters are the same as in Fig. 2.

**Fig. 5 – Bow-tie can evolve for a nonlinear input-output relation too, if the input can me more compactly represented with no effect on the output.** We show simulation results of a simple nonlinear problem mimicking a 4-pixel retina. **(A) Problem definition**: The retina has four inputs (one for each pixel, that can be either black or white), four outputs and two internal processing layers. The retina is evolved so that its outputs detect whether there is (i) an object on the left side (at least one pixel in the left column is black), (ii) on the right side (at least one pixel in the right column is black), (iii) left AND right objects, (iv) left OR right objects, correspondingly. Inset: in contrast to previous problems, here each node performs a nonlinear transformation of the sum of weighted inputs: $u^{(l+1)} = f(A^{(l)}u^{(l)} - T^{(l+1)})$, where $A^{(l)}$ and $T^{(l)}$ are the weight matrix and set of thresholds in the $l$-th layer. **(B) Typical example of simulation results**. Apparently, two bits of information are sufficient to fully describe the four required outputs in this model. Indeed, the network evolved so that it has only two active nodes in the second layer (red circles).

**Fig. 6 – A bow-tie architecture is obtained under a broad range of evolutionary parameters.** We tested the existence and width of bow-ties under a broad range of parameter values. We illustrate here the mean and standard deviation of the bow-tie width for various values of mutation rate, mutation size, population size and selection intensity. Bow-ties were obtained in all cases. The width of the bow-tie showed little sensitivity to the parameter values. Each point is based on 50 independent repeats of the simulation. Parameter values tested: population size = [50, 100, 250, 500]; mutation size = [0.01, 0.05, 0.1, 0.2, 0.5, 1]; mutation rate = [1, 0.25, 0.1, 0.05]/$LD^2$, tournament size $s$ = [2, 4, 6, 8].

**Table 1: examples for networks having bow-tie (hourglass) architecture**

| network | Input | intermediate | output | Refs. |
|---|---|---|---|---|
| Metabolic network | Nutrients | Precursor metabolites (among them G6P, F6P, PEP, PYR, AKG, ACCOA) | Complex macromolecules (proteins, fatty acids, carbohydrates) | (1, 7) |
| Developmental gene regulatory network | Patterning genes (e.g. Hox genes, *wingless*, *EGF-R*, *hedgehog*, *Notch*) | 'input-output' or 'selector' gene, e.g. *shavenbaby* or *scute* in Drosophila | Developmental program of epidermis formation. | (5) |
| Innate immune response | >1000 microbial molecules | 10 Toll-like receptors (TLRs), 4 TIR (Toll/interleukin-1 receptor homologous region) adaptors and 2 protein kinases (19) MyD88 gene (18) | Genes responsive to NF-kB and STAT1 (>500), secondary and tertiary events (>1000) | (18, 19) |
| Immune system | Environmental stimuli (bacteria, viruses, toxins) | Immature dendritic cells, naïve CD4+ T-cells | Various cytokines and antibody releases (B-cells, CD8+ T-cells, macrophages, etc.) | (4, 20) |
| Signaling networks | receptors | cAMP, calcium | Genes | (16, 17) |
| Human visual system | $10^8$ photoreceptors in the eye | Optic nerve (axons of $10^6$ ganglion cells) | Image features: color, pattern, depth, movement, etc. | (14, 15) |
| Multi-layered (deep) neural network used for clustering and dimensionality reduction | Real data is fed to upper layers ("encoder"). | "bottleneck" | Reconstructed data which is the product of the lower layers ("decoder"). | (71) |
| Internet protocols | Physical layer (coaxial cable, optic fiber, CDMA, TDMA) | Network layer (IPv4/IPv6 protocol) | Application layer (Thunderbird, Firefox, skype) | (3) |

**Figures:**

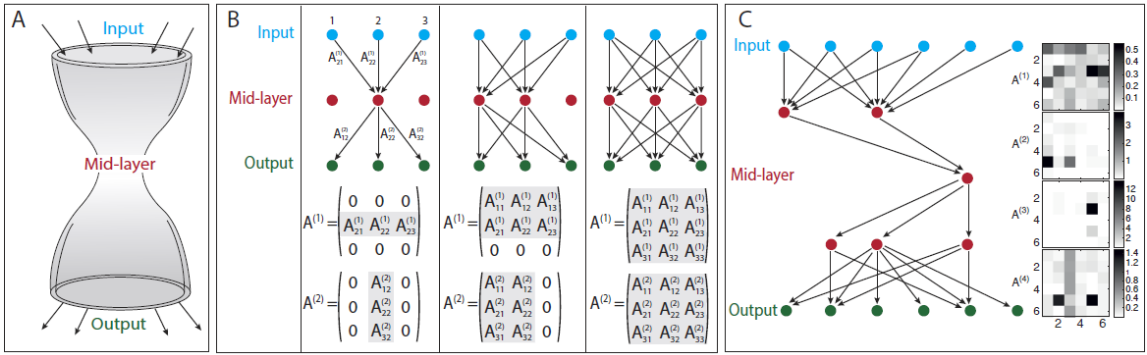

**Fig. 1**

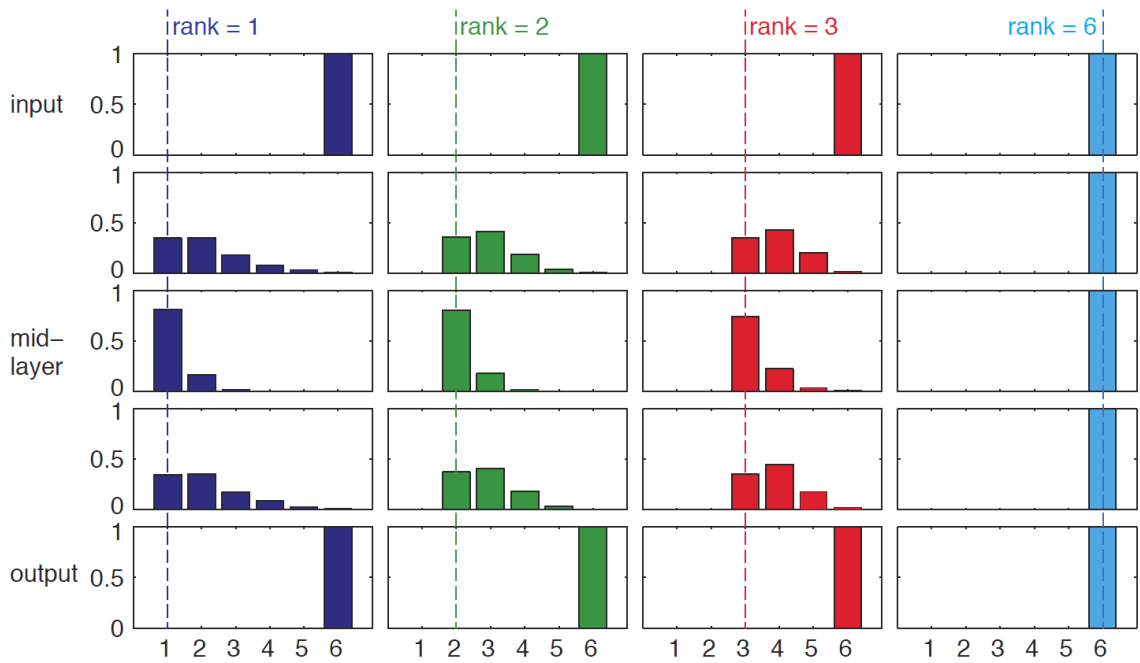

**Fig. 2**

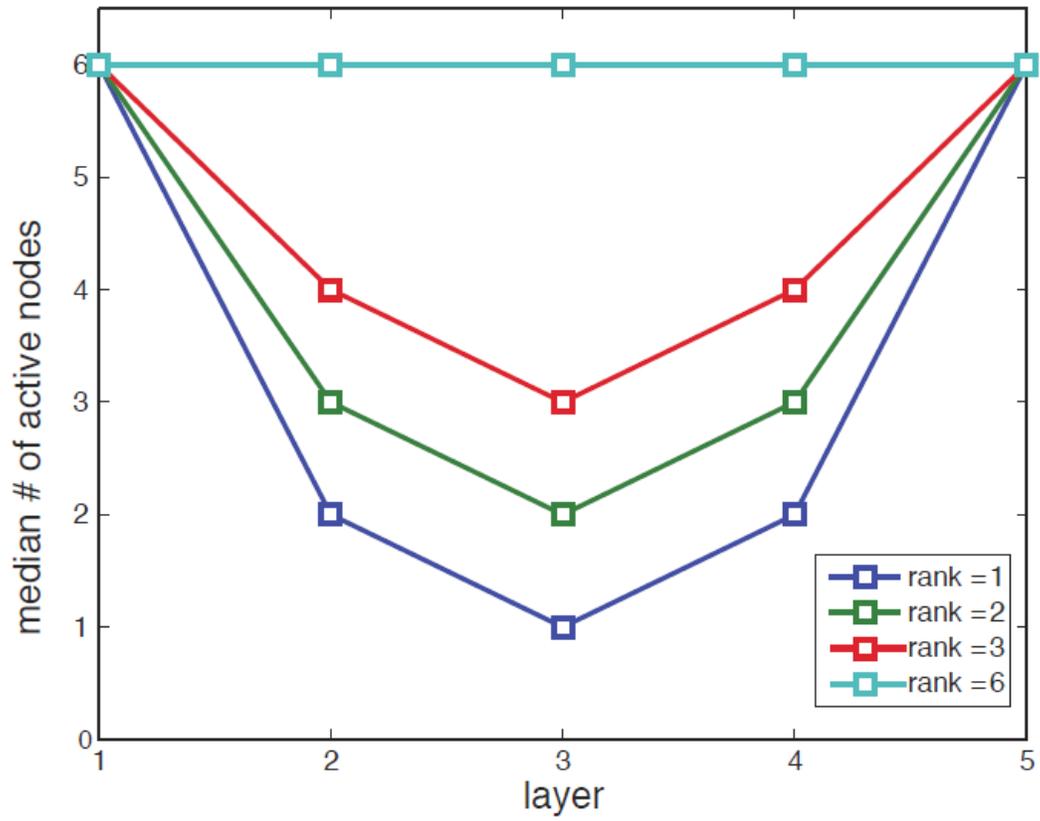

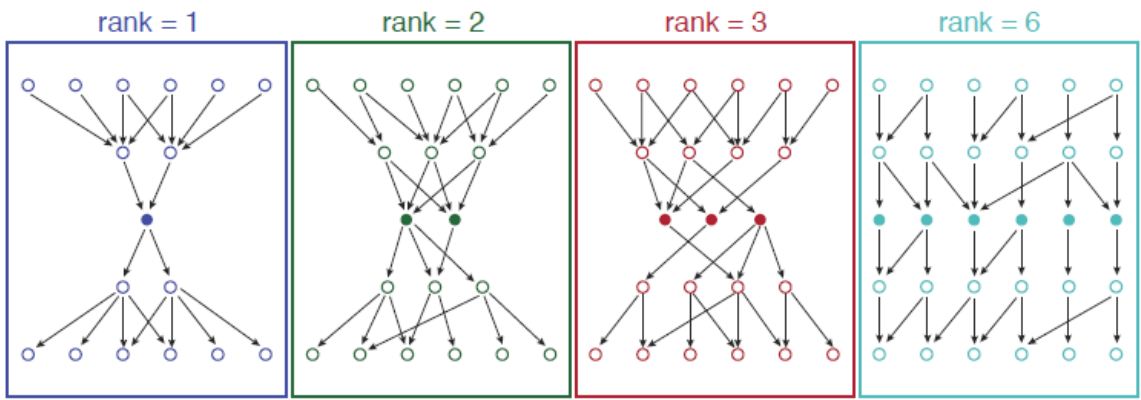

**Fig. 3**

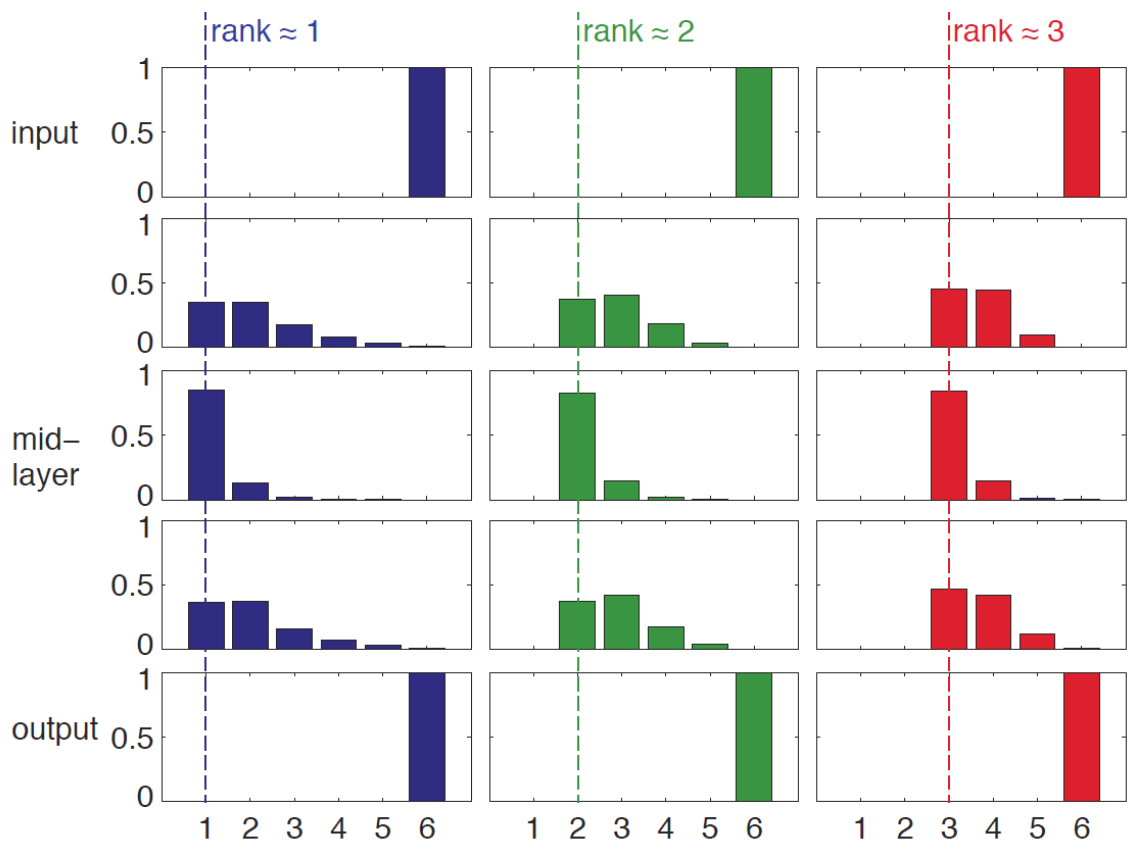

**Fig. 4**

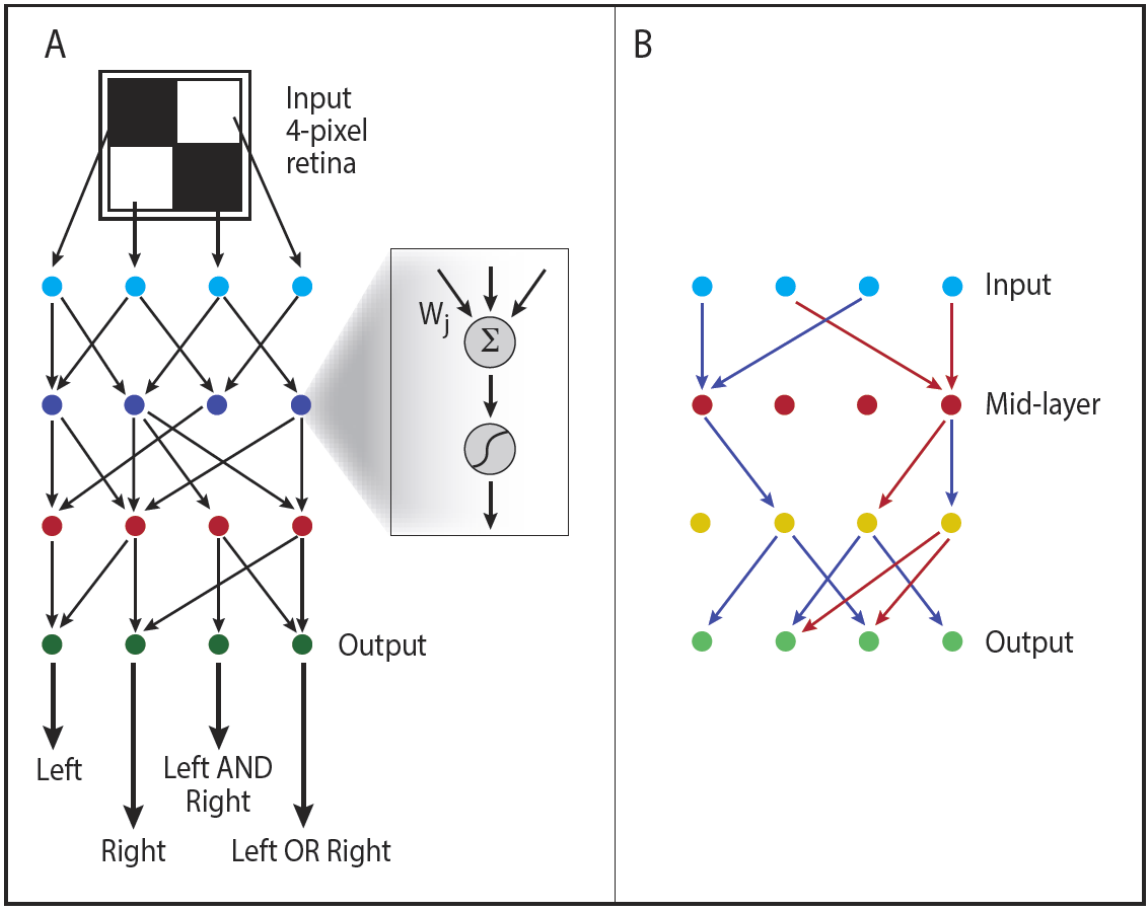

**Fig. 5**

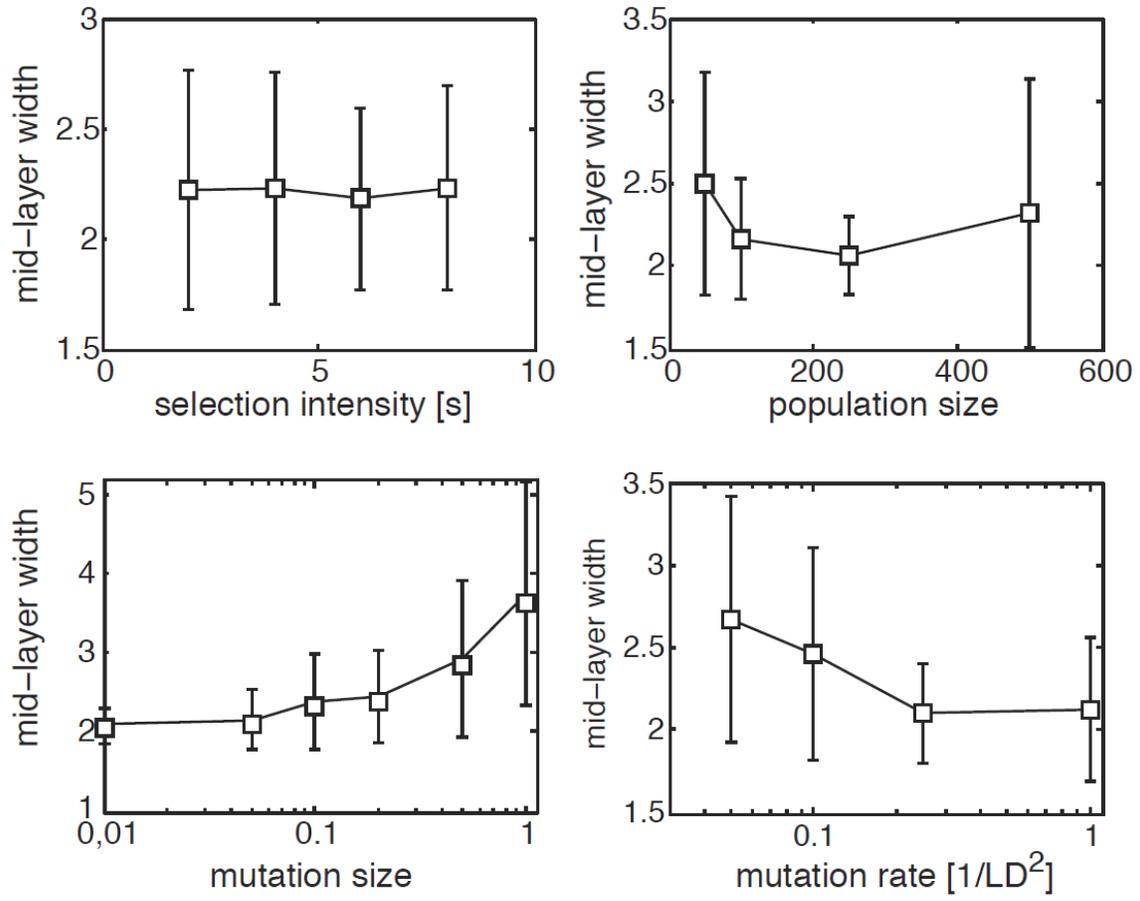

**Fig. 6**

# Evolution of bow-tie architectures in biology

Tamar Friedlander, Avraham E. Mayo, Tsvi Tlusty and Uri Alon

**Supporting Information**

## Table of Contents

# 1. Sensitivity to parameters

Evolutionary simulations results depend on a number of key parameters. Here we tested the sensitivity of our simulation results: the emergence of bow-tie structures with narrow intermediate layer, best population fitness reached and fitness temporal trajectories and to the following simulation parameters:

1. Selection intensity (size of tournament)
2. Mutation size
3. Mutation rate
4. Population size.

All parameters affect both the speed of fitness convergence and the final fitness value achieved to some extent. Bow-tie width was found insensitive to the selection intensity and weakly sensitive to population size. It showed some dependence on mutation size and rate. Yet, bow-ties did evolve under a broad range of values for all these parameters.

Parameters for all runs: $L=4, D=6$ (namely a network of 5 layers with 6 nodes in each), goal:

$$G = \begin{bmatrix} 10 & 10 & 10 & & & \\ 10 & 10 & 10 & & 0 & \\ 10 & 10 & 10 & & & \\ & & & 10 & 10 & 10 \\ & 0 & & 10 & 10 & 10 \\ & & & 10 & 10 & 10 \end{bmatrix}$$ with rank 2.

Here we detail the values tested for these parameters and the default value used otherwise.

Mutation values were drawn from a normal distribution with $N(1,\sigma)$. Default mutation size (except for mutation size run) used was $\sigma = 0.1$. In runs testing mutation size we tested the values $\sigma = 0.01, 0.05, 0.1, 0.2, 0.5, 1$.

Default mutation rate (per matrix entry) used $R_0/5$, where $R_0 = 1/LD^2$. In runs testing mutation rate we tested the values: $R_0$, $R_0/4$, $R_0/10$, $R_0/20$.

Default tournament size (except for tournament size runs) $s = 4$. In runs testing selection intensity we tested the values $s = 2, 4, 8$. This parameter determines the selection intensity.

Default population size was 100. In runs testing population size effects we tested the values 50, 100, 250, 500.

In the following we illustrate examples of temporal fitness trajectories (population mean) and histograms of layer width with different parameter values.

**Selection intensity**

Selection intensity had a negligible effect on the bow-tie structure, and was mainly affecting the speed of convergence to better fitness (see Figs. S1, S2). The stronger is the selection (larger tournament size), the closer is the fitness to the optimum at a given number of generations. Note that the figures show the absolute value of fitness (which is the distance from the optimum) – so the curves are monotonously decreasing.

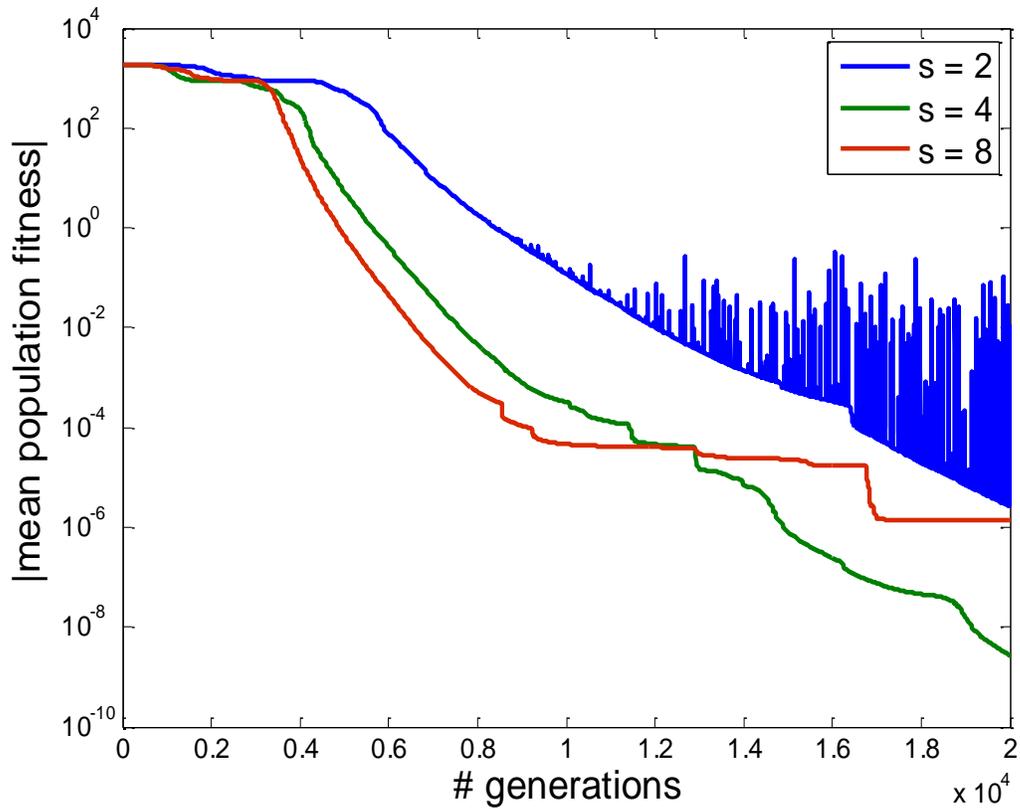

**Fig S1: The Stronger the selection, the better is the fitness achieved at a given number of generations.** We show here examples of fitness temporal trajectories with different tournament size parameter (determines selection intensity). Other parameters are the same in all three simulations.

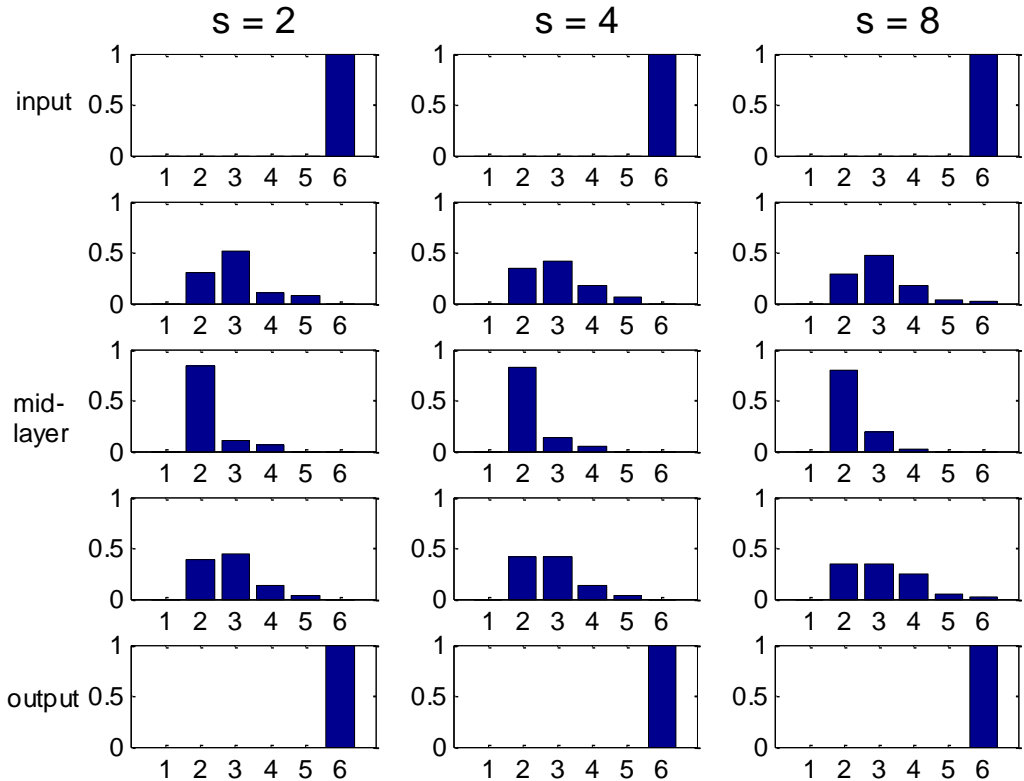

**Fig S2: The emergences of narrow bow-tie (and in general layer width) are insensitive to selection intensity.** Simulation results showing layer width statistics for different selection intensity values, where all other parameters remain intact. Statistics is based on 100 different runs for each selection intensity value.

**Mutation size and rate**

Examining the effect of mutation size and rate on temporal fitness trajectories, one notices two phases: fast convergence at the first phase followed by slow convergence at the second phase. The larger is the mutation size / rate the faster is the fitness convergence in the initial phase, but it is then "stuck" at a fitness value which is further from the optimum in the second phase. This can be thought of as a "noise floor" defined by the mutations (see Figs. S3 and S5).

Generally, all values of mutation size and rate tested show a bow-tie architecture. The smaller is the mutation size (the higher the mutation rate) , the closer is the width of the bow-to the minimal size possible. In this case the rank of the goal was 2, which defines the minimum – see Figs. S4, S6.

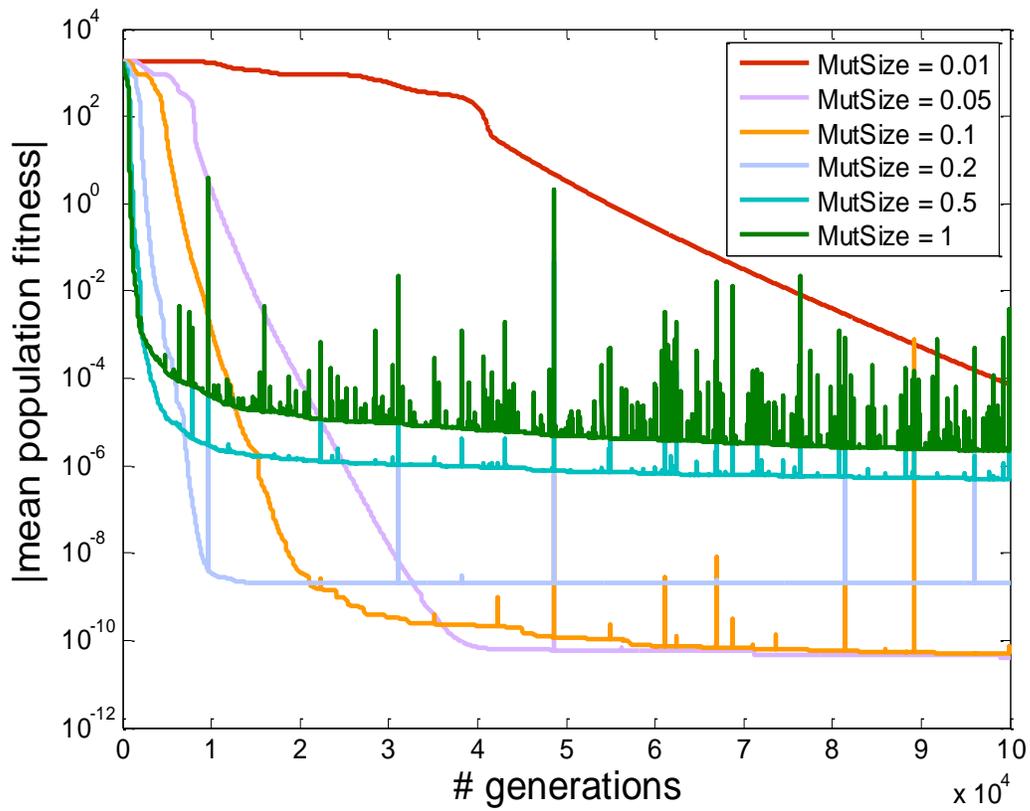

**Fig S3: Mutation size affects both the speed of fitness convergence at the initial phase and the finite fitness value reached at the second phase.** The figures shows examples of fitness temporal trajectories with different values of mutation size (other parameters are the same). The smaller is the mutation size, the slower is the convergence at the first phase, but the final fitness reached is higher. Note that the red curve (mutation size = 0.01) did not converge in the time tested here.

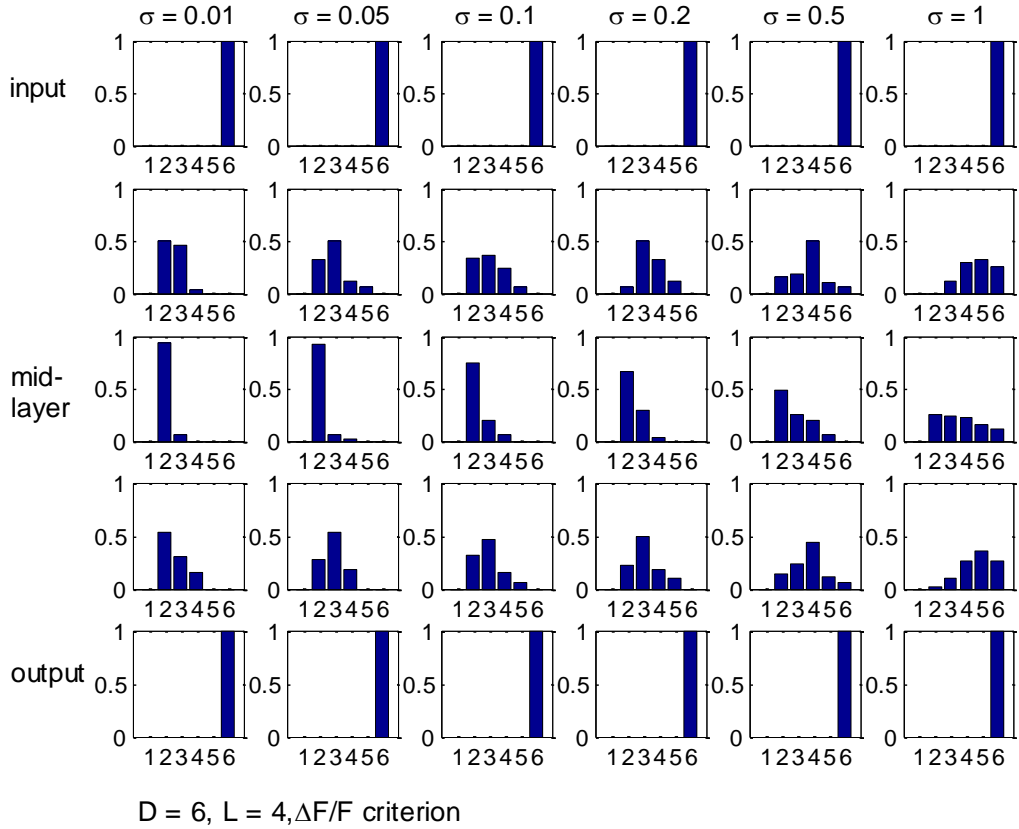

D = 6, L = 4, ΔF/F criterion

**Fig S4: The smaller is the mutation size, the narrower is the waist layer.** Simulation results showing layer width statistics for different mutation size values, where all other parameters remain intact. Statistics is based on 50 different runs for each mutation size value. Runs lasted 100,000 generations each regardless of the fitness value. Typically, fitness values reached were very close to the optimum (order of $10^{-4}$-$10^{-6}$ from the optimum).

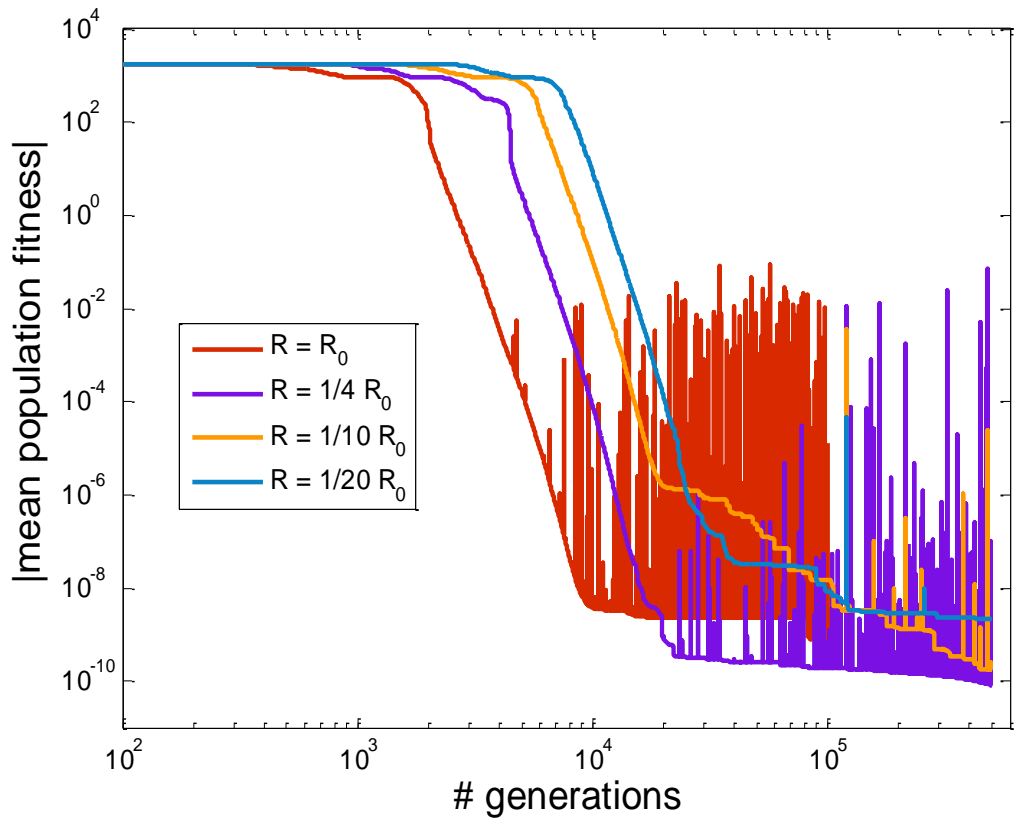

**Fig S5: Mutation rate affects both the speed of fitness convergence at the initial phase and the finite fitness value reached at the second phase.** The figure shows examples of fitness temporal trajectories with different values of mutation rate, where $R_0 = 1/LD^2$. Other simulation parameters are the same for all curves.

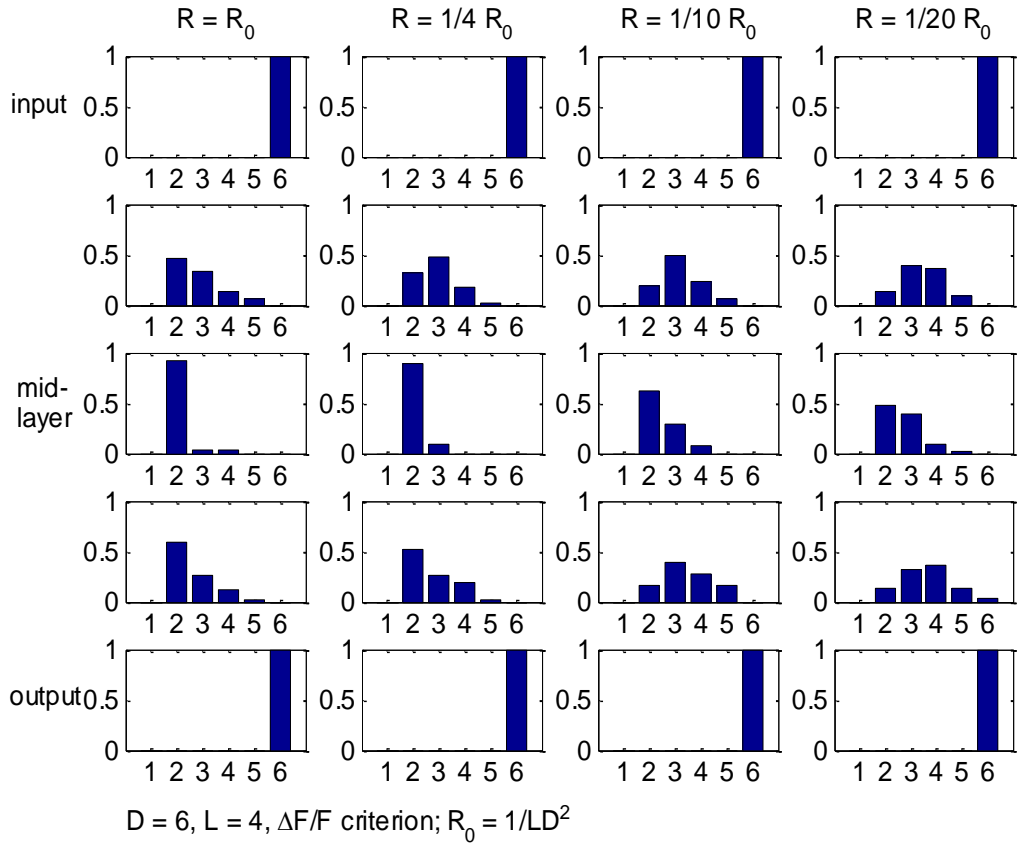

D = 6, L = 4, ΔF/F criterion; $R_0 = 1/LD^2$

**Fig S6: The higher is the mutation rate, the narrower is the waist layer (for equal number of generations).** Simulation results showing layer width statistics for different mutation rate values, where all other parameters remain intact. Statistics is based on 50 different runs for each mutation rate value. Length of runs was 100,000 generations for the highest mutation rate ($R=R_0$) and 500,000 generations for the other 3 rates. Fitness values reached within this time were very close to the optimum (within $10^{-8}$-$10^{-16}$ from the optimum for the highest mutation rate and within $10^{-6}$-$10^{-10}$ for the lowest mutation rate).

**population size**

Population size also showed different effects on the two phases of fitness trajectories, in similarity to mutation size and rate. The larger is the population, the slower is the fitness convergence at the first phase. The second phase was highly variable and at different repeats of the simulation different population sizes reached the best fitness (Fig. S7).

Bow-tie width showed a weak dependence on population size. In the smallest population size tested (N=50) the bow-tie obtained was not as narrow as the goal rank allows. However, between the other 3 population size values tested (100, 250, 500) there only little difference (Fig. S8).

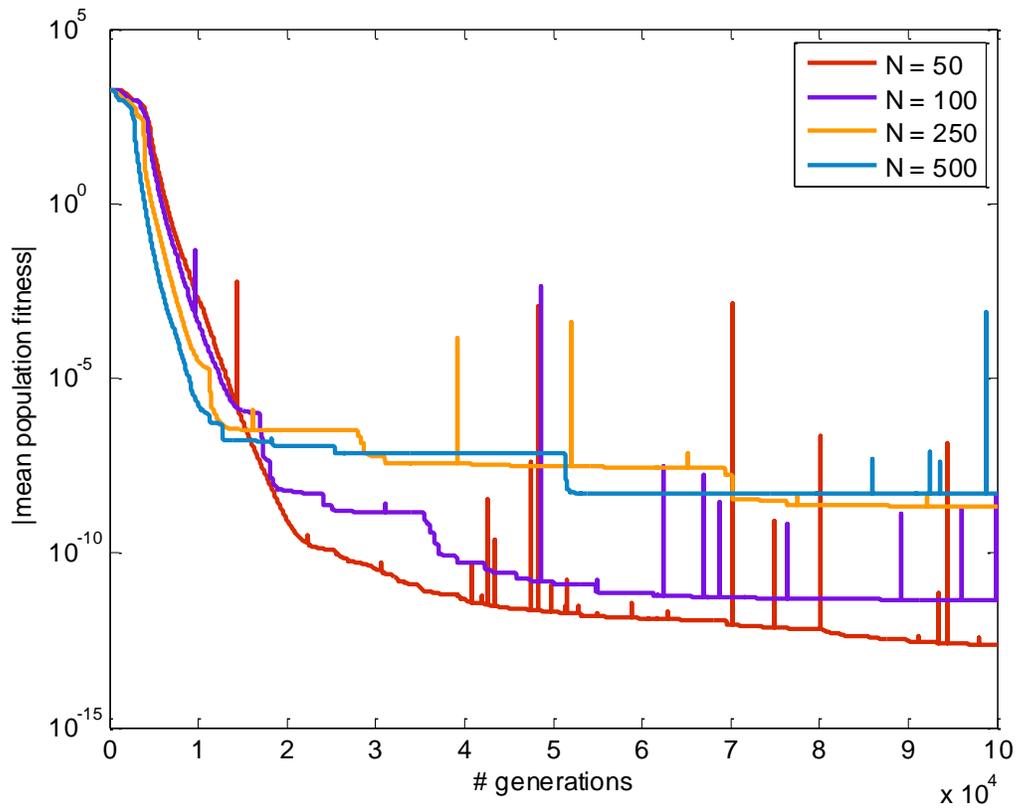

**Fig S7: Population size affects the speed of fitness convergence at the initial phase – the larger is the population size, the faster the convergence.** The finite fitness value reached is highly variable, and at different repeats of the simulation different population sizes reach the best fitness. Here we show examples of fitness temporal trajectories with different values of population size.

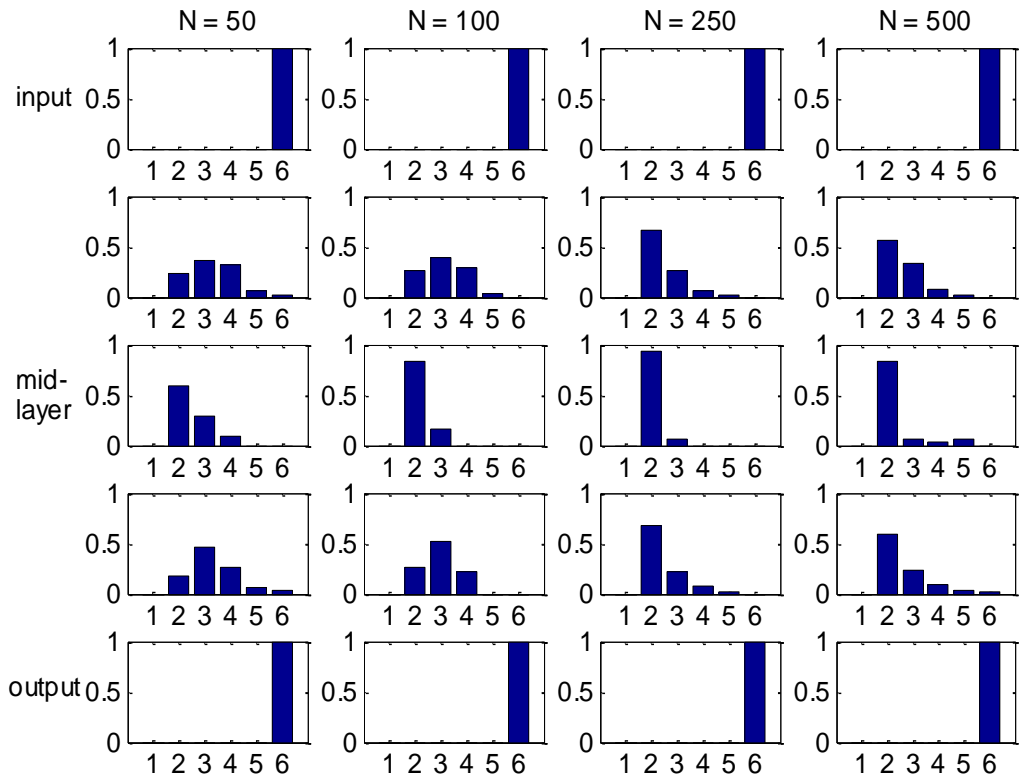

**Fig S8: The larger is the population size, the narrower is the waist layer (for equal number of generations).** Simulation results showing layer width statistics for different population size values, where all other parameters remain intact. Statistics is based on 50 different runs for each population size value. Number of generations was 100,000 at each run.

## 2. The emergence of bow-tie is insensitive to the internal goal structure (as long as the rank remains intact)

We compared the network structure obtained with different goal matrices that have the same rank but different internal structure. In particular we were interested whether the number of zeroes can affect the structure obtained. We found that the steepness of active node decline towards the waist can be affected. However the width and location of the waist layer are insensitive to the exact number of zeroes in the goal - See Fig. S9 for illustration. Matrices used as goals were:

$$G_8 = \begin{bmatrix} 10 & 10 & 10 & 10 & 10 & 10 & 10 & 0 \\ 10 & 10 & 10 & 10 & 10 & 10 & 10 & 0 \\ 10 & 10 & 10 & 10 & 10 & 10 & 0 & 10 \\ 10 & 10 & 10 & 10 & 10 & 10 & 0 & 10 \\ 10 & 10 & 10 & 10 & 10 & 0 & 10 & 10 \\ 10 & 10 & 10 & 10 & 10 & 0 & 10 & 10 \\ 10 & 10 & 10 & 10 & 0 & 10 & 10 & 10 \\ 10 & 10 & 10 & 10 & 0 & 10 & 10 & 10 \end{bmatrix},$$

$$G_{16} = \begin{bmatrix} 0 & 0 & 10 & 10 & 10 & 10 & 10 & 10 \\ 0 & 0 & 10 & 10 & 10 & 10 & 10 & 10 \\ 10 & 10 & 0 & 0 & 10 & 10 & 10 & 10 \\ 10 & 10 & 0 & 0 & 10 & 10 & 10 & 10 \\ 10 & 10 & 10 & 10 & 0 & 0 & 10 & 10 \\ 10 & 10 & 10 & 10 & 0 & 0 & 10 & 10 \\ 10 & 10 & 10 & 10 & 10 & 10 & 0 & 0 \\ 10 & 10 & 10 & 10 & 10 & 10 & 0 & 0 \end{bmatrix},$$

$$G_{32} = \begin{bmatrix} 10 & 0 & 10 & 0 & 10 & 0 & 10 & 0 \\ 10 & 0 & 10 & 0 & 10 & 0 & 10 & 0 \\ 10 & 10 & 0 & 0 & 10 & 10 & 0 & 0 \\ 10 & 10 & 0 & 0 & 10 & 10 & 0 & 0 \\ 10 & 10 & 10 & 10 & 0 & 0 & 0 & 0 \\ 10 & 10 & 10 & 10 & 0 & 0 & 0 & 0 \\ 0 & 10 & 0 & 10 & 0 & 10 & 0 & 10 \\ 0 & 10 & 0 & 10 & 0 & 10 & 0 & 10 \end{bmatrix},$$

$$G_{48} = \begin{bmatrix} 10 & 10 & 0 & 0 & 0 & 0 & 0 & 0 \\ 10 & 10 & 0 & 0 & 0 & 0 & 0 & 0 \\ 0 & 0 & 10 & 10 & 0 & 0 & 0 & 0 \\ 0 & 0 & 10 & 10 & 0 & 0 & 0 & 0 \\ 0 & 0 & 0 & 0 & 10 & 10 & 0 & 0 \\ 0 & 0 & 0 & 0 & 10 & 10 & 0 & 0 \\ 0 & 0 & 0 & 0 & 0 & 0 & 10 & 10 \\ 0 & 0 & 0 & 0 & 0 & 0 & 10 & 10 \end{bmatrix}.$$

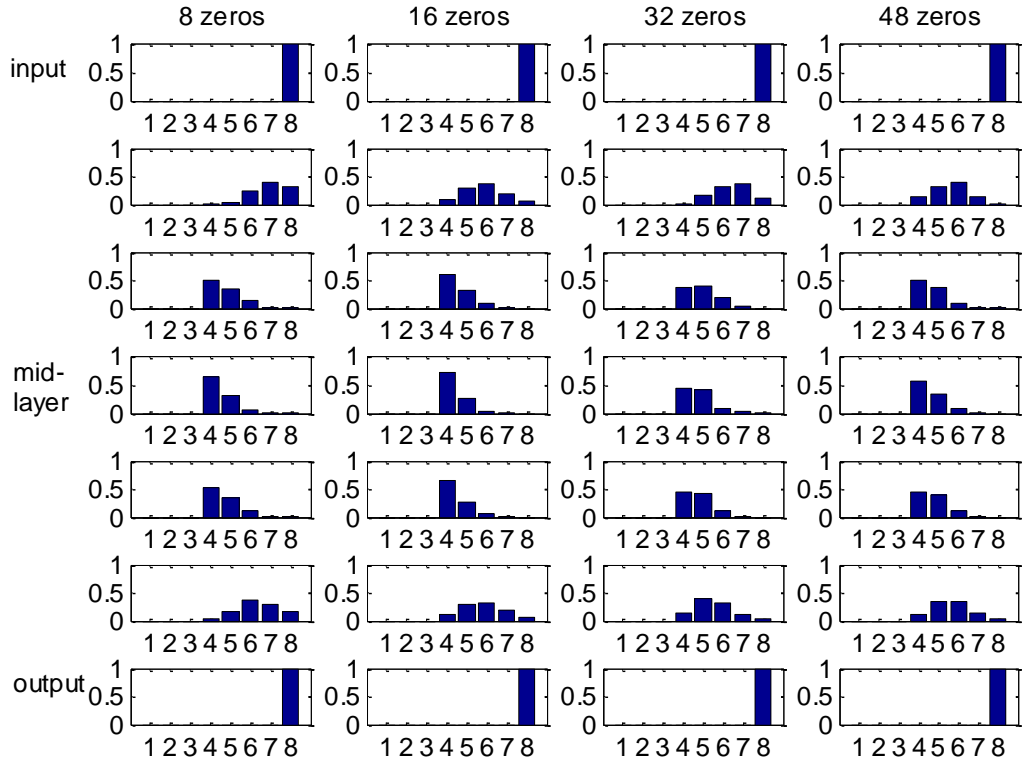

**Fig S9: Layer width is insensitive to the internal goal structure, as long as the rank doesn't change.** Simulation results showing layer width statistics for goals that all have rank 4, but different number of zero terms, where all other parameters remain intact. Statistics is based on 500 different runs for each goal structure. Only runs that converged to fitness value within 0.01 from the optimum were considered (# runs analyzed = {488   498   390   497}, correspondingly). Simulation parameters: D=6, L=8. Simulations were run for 100,000 generations.

## 3. Sum-mutations are less likely to lead to narrow bow-tie structures compared to product-mutations

We show for control the results of a similar evolutionary simulation in which sum-mutations were used instead of product mutations (and all other simulation parameters are intact). The sum-rule is the commonly used addition of a normally distributed random number to a randomly chosen element of the matrices, which represents a mutation in the intensity of a single interaction between network components, $A_{ij} \to A_{ij} + \mathrm{N}(0,\sigma)$.

In contrast, under product-rule mutations, an element of the matrix is multiplied by a random number drawn from a normal distribution with mean 1: $A_{ij} \to A_{ij} \cdot \mathrm{N}(1,\sigma)$. For a comprehensive comparison of sum and product mutation rules the reader is referred to an earlier work (24). Here we compare the distributions of connection intensities obtained under these two mutational schemes in our problem (Fig. S10). Product-mutations result in a distribution that has broader tails towards both high and low-values, which explains why more connections vanish and more nodes become inactive under product-mutations.

We analyze the number of nodes per network layer over repeats of the simulation. Ranks of the goals tested were 1 and 2. While under product-mutations a waist of width 1 evolved in the middle layer in most runs, here under sum-mutations its width was typically 3-4 (see Fig S11 and Table S1 below for summary of simulation results).

Similarly, in the nonlinear retinal problem network structure obtained under sum-mutations did not have as narrow a waist as with product-mutations (see Fig S12 for statistics of network structures under the two mutational schemes).

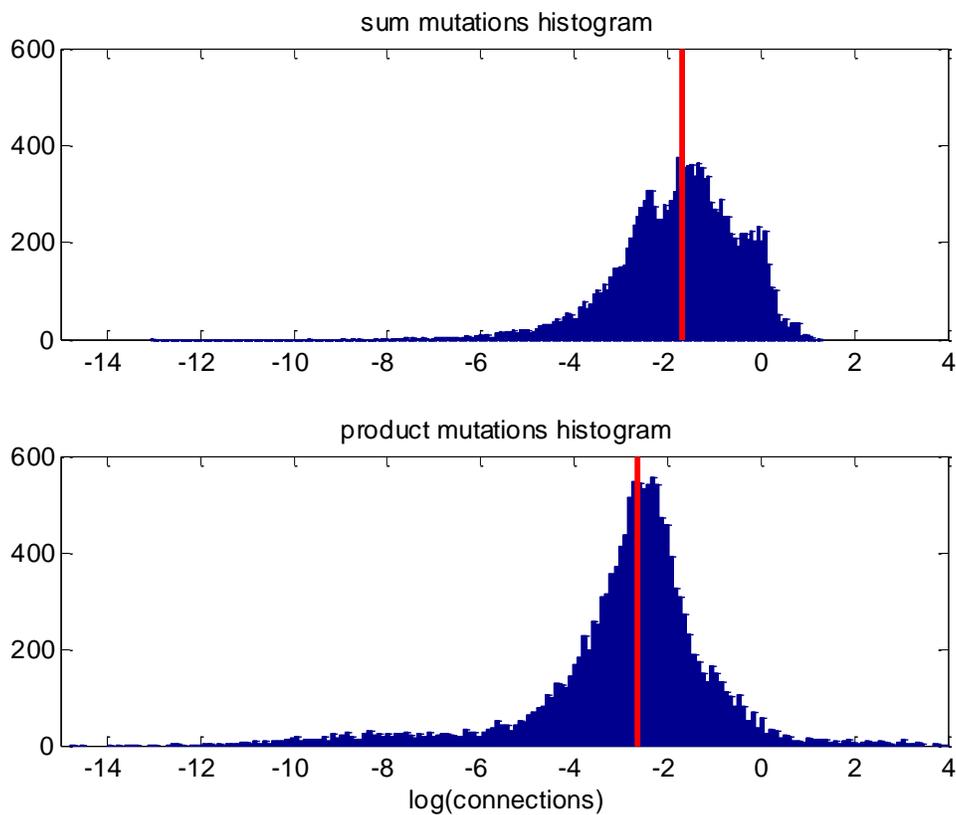

**Fig S10: Sum mutation connection intensities exhibit a much narrower distribution compared to product mutations** – note that the x axis is the logarithm of the absolute value of connection intensities. Histograms were produced by taking all matrix terms from 100 independent runs for each mutation type. Parameters: D=6, L=4, which results in 14,400 terms in each case. Red line designates the log of the median for each distribution.

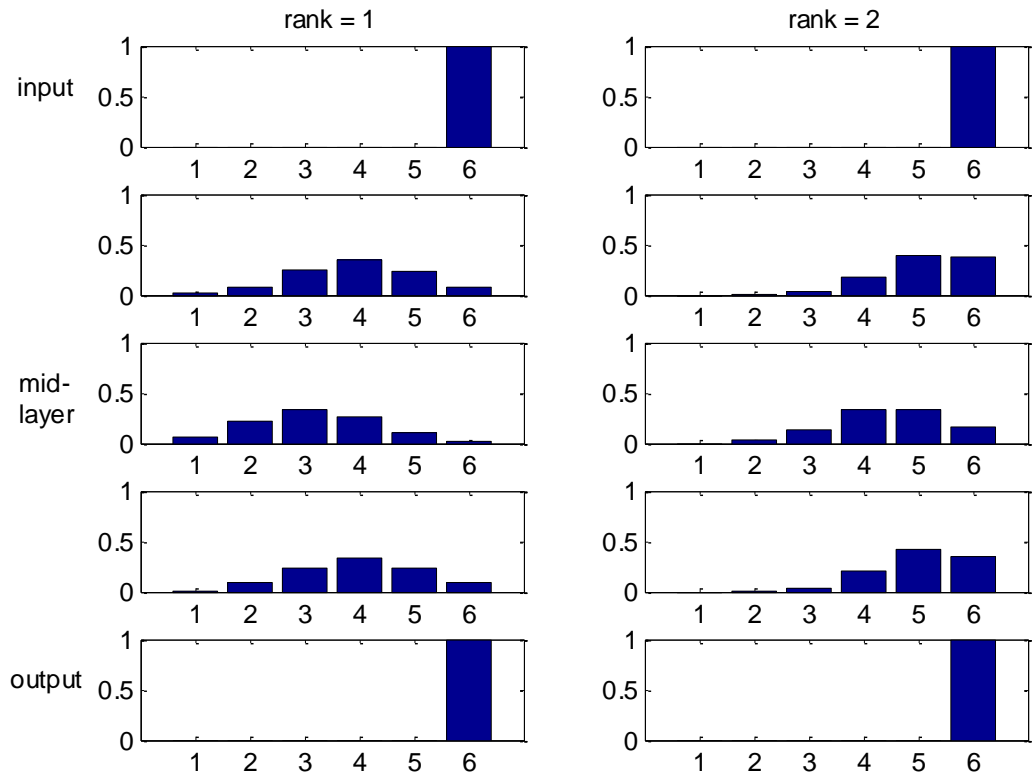

**Fig S11: Sum-rule mutations do not lead to a narrow bow-tie for rank-deficient goals, as product-mutations do.** We show here layer width statistics for goals of rank 1 (left column) and 2 (right column). Statistics is based on 360 repeats for each goal. Simulation was run for 100,000 generations each time. All runs converged to fitness value within 0.01 from the optimum.

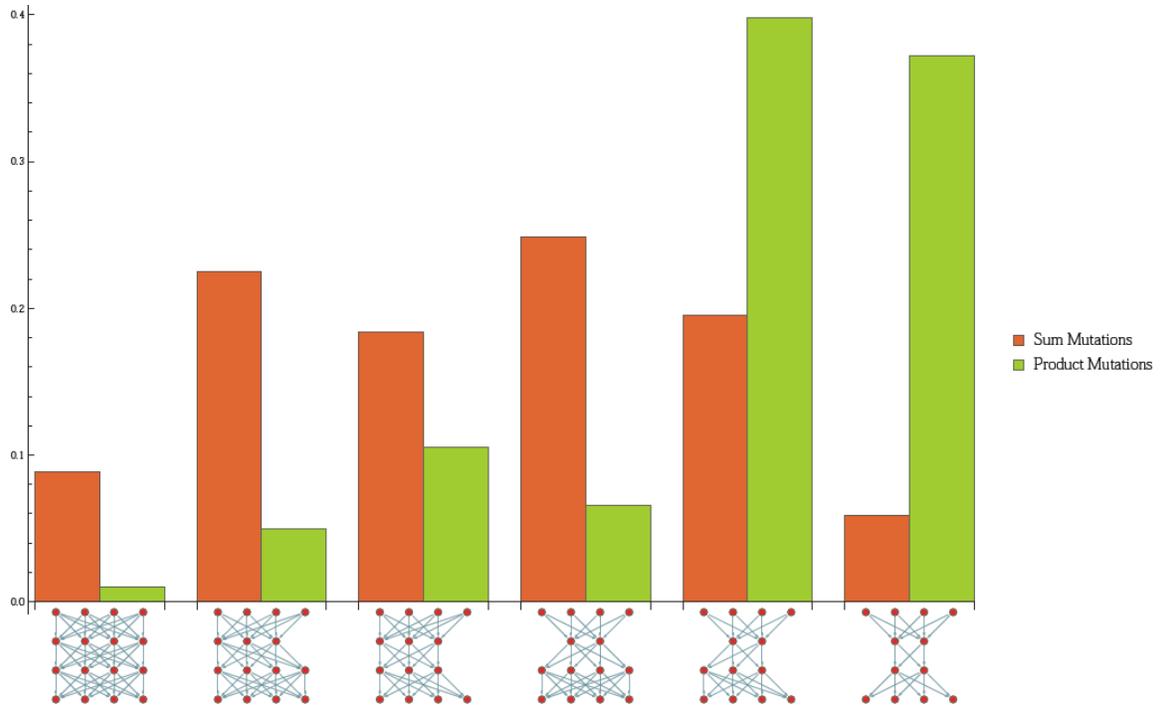

**Fig S12: In the nonlinear problem too, sum-mutations are less likely to lead to a narrow bow-tie under rank-deficient goal compared to product mutations.** Here we show a histogram of possible network structures obtained in repeated runs in the nonlinear problem with either sum or product-mutations.

# 4. Fraction of runs that did not converge to a bow-tie with narrow layer that equals the goal rank

| Linear problem - $D=6$; $L=4$, product mutations, $\sigma=0.1$ | | |
|---|---|---|
| rank | number of runs (*) | Fraction of runs with bow-tie > rank |
| 1 | 2996 | 0.19 |
| 2 | 2986 | 0.20 |
| 3 | 1500 | 0.26 |
| 6 | 199 | 0 (**) |
| Linear problem - $D=6$; $L=4$, product mutations, $\sigma=0.5$ (2% sign change) | | |
| rank | | |
| 1 | 425 | 0.37 |
| 2 | 425 | 0.48 |
| 3 | 425 | 0.51 |
| Linear problem - $D=6$; $L=4$, sum mutations | | |
| rank | number of runs(*) | Fraction of runs with bow-tie > rank |
| 1 | 1498 | 0.94 |
| 2 | 1484 | 0.97 |
| Nonlinear retina problem - $D=4$; $L=3$, product mutations | | |
| 2 | 505 | 0.16 |
| Nonlinear retina problem - $D=4$; $L=3$, sum mutations | | |
| 2 | 544 | 0.50 |

Table S1: Fraction of runs that did not reach a bow-tie configuration with a layer as narrow as the goal rank under either product or sum mutations for both linear and nonlinear models.

(*) - Runs were considered in the analysis only if they reached within 0.01 (or less) than the optimal fitness possible.

(**) - In the case of rank equal to 6 all network configurations must have 6 nodes at each row. Since this is also the dimension of matrices, it is not possible to have a number of nodes which is larger than the rank.

# 5. A bow-tie evolves even if the product-mutations can change interaction sign

We tested the evolutionary simulation with product mutations that were drawn from a broader normal distribution N(1,0.5) that has probability of 2% to change the sign of the mutated element. Here too bow-ties evolved. For detailed statistics of simulation results with goals having ranks 1, 2, 3 see Fig S13 below.

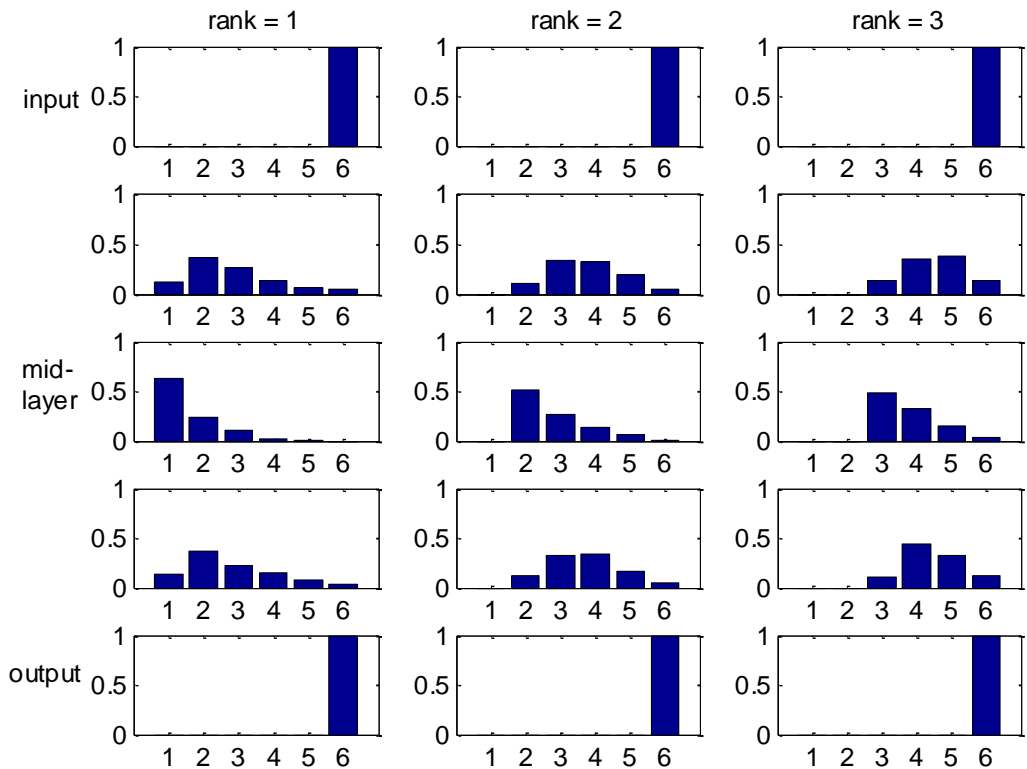

**Fig S13: Bow-ties evolve even when the mutations have a significant probability to change the interaction sign.** Here we drew product-mutations from a broader distribution N(1,0.5), from which 2% of the mutations would change the sign of the mutated element. We show here layer width statistics for goals of ranks 1, 2 and 3. Other simulation parameters are intact. Specifically the per-term mutation rate is 0.0014. Statistics is based on 425 independent runs for each rank. Runs were stopped when the fitness reached within 0.01 of the optimum. All runs reached this limit.

# 6. Bow-tie dependence on noise level added to the goal

We tested the ability of the evolutionary mechanism to filter out noise added to the goal, by varying the noise level. In Fig S14 we illustrate the width of the bow-tie obtained vs. the noise level added to a goal whose "clean rank" was 1. See noise level definition in the Methods Section (main text).

The noise was drawn from normal distributions with different standard deviation values (0.1, 0.2, 0.5, 1). Noise terms were drawn independently for each repeat of the simulation, such that the evolutionary goal was slightly different for each run (but didn't change during the course of the run). Results are based on 780 independent runs for each noise level.

The "clean goal" was: $G_{clean} = \begin{bmatrix} 10 & 10 & 10 & 10 & 10 & 10 \\ 10 & 10 & 10 & 10 & 10 & 10 \\ 10 & 10 & 10 & 10 & 10 & 10 \\ 10 & 10 & 10 & 10 & 10 & 10 \\ 10 & 10 & 10 & 10 & 10 & 10 \\ 10 & 10 & 10 & 10 & 10 & 10 \end{bmatrix}$.

An example of a noisy goal (lowest noise level):

$G_{noisy} = \begin{bmatrix} 9.9590 & 10.1611 & 10.0919 & 10.0580 & 9.8794 & 10.1750 \\ 9.9630 & 9.9266 & 10.0948 & 10.0780 & 10.0377 & 9.8305 \\ 9.9102 & 9.9060 & 10.0166 & 10.0859 & 9.8447 & 10.0698 \\ 9.9437 & 10.0309 & 9.9075 & 9.8582 & 9.8845 & 9.9820 \\ 9.8180 & 10.1331 & 9.8766 & 10.0667 & 9.8966 & 10.0801 \\ 10.0806 & 10.0267 & 10.0364 & 10.0566 & 9.9908 & 10.2014 \end{bmatrix}$.

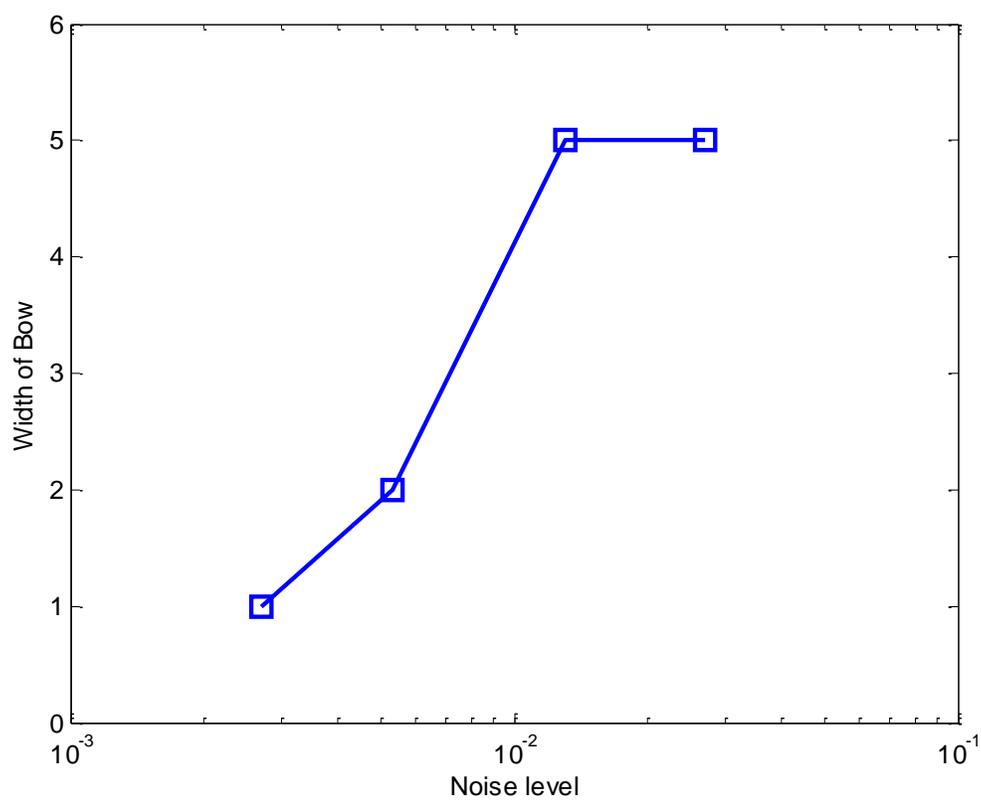

**Fig S14: Width of both tie increases with noise level added to the goal. The evolutionary mechanism is able to filter out noise to a certain extent, and expose the clean rank of the goal, however when noise becomes too large the evolutionary mechanism fails in filtering. Here the clean rank of the goal was 1 and the noise was drawn from normal distributions with increasing standard deviations. Noise level was defined as the absolute value of the difference between noisy and clean goal norms divided by the clean goal norm (see Methods Section).**

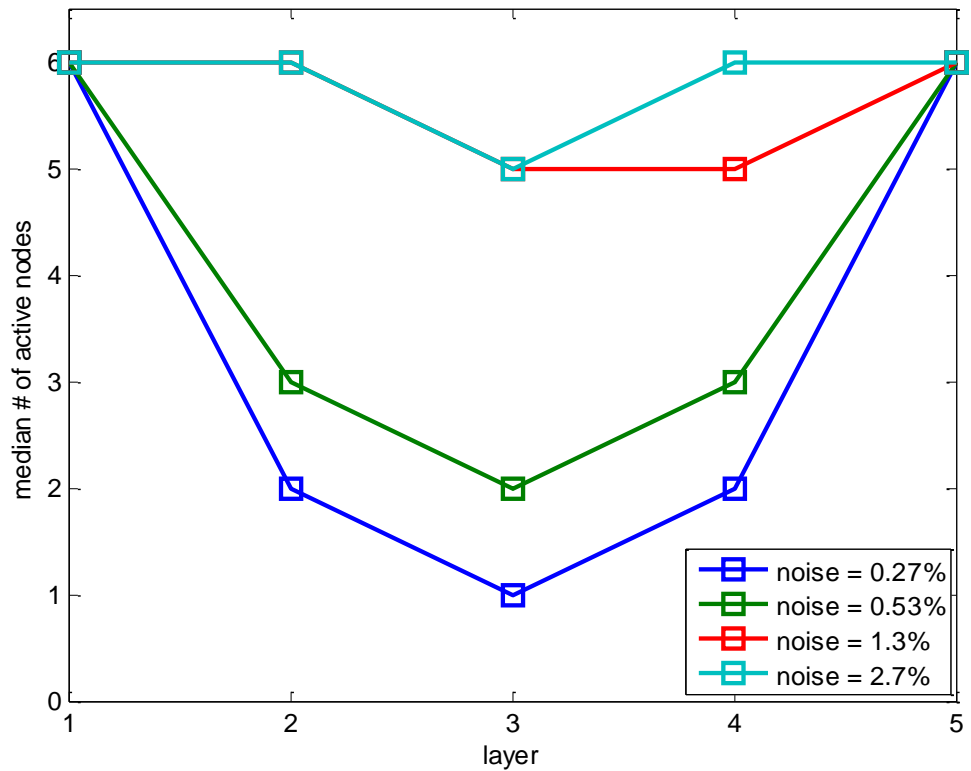

**Fig S15: Median number of active nodes per layer for varying noise levels added to the goal. Results refer to the same simulations as in the previous figure.**